\documentstyle[epsfig]{elsart}
\begin{document}
\message{\the\hsize}
\def\sim{\Sigma^-}
\def\pim{\pi^-}
\def\pip{\pi^+}
\def\pbp{\bar{\textnormal{p}}\textnormal{p}}
\def\tot{\sigma_{\rm tot}}
\def\totsimp{\sigma_{\rm tot}(\Sigma^- \rm p )}
\def\totpp{\sigma_{\rm tot}(\rm pp)}
\def\totpimp{\sigma_{\rm tot}(\pi^- \rm p )}
\def\totkmip{\sigma_{\rm tot}(\rm K^- \rm p )}
\def\totpimbe{\sigma_{\rm tot}(\pi^- \rm Be )}
\def\totpipbe{\sigma_{\rm tot}(\pi^+ \rm Be )}
\def\totpimc{\sigma_{\rm tot}(\pi^- \rm C )}
\def\totpipc{\sigma_{\rm tot}(\pi^+ \rm C )}
\def\totpimcu{\sigma_{\rm tot}(\pi^- \rm Cu )}
\def\totsimbe{\sigma_{\rm tot}(\Sigma^- \rm Be )}
\def\totsimc{\sigma_{\rm tot}(\Sigma^- \rm C )}
\def\totsimcu{\sigma_{\rm tot}(\Sigma^- \rm Cu )}
\def\totsimpol{\sigma_{\rm tot}(\Sigma^- \rm CH_2 )}
\def\totpimpol{\sigma_{\rm tot}(\pi^- \rm CH_2 )}
\def\totpbe{\sigma_{\rm tot}(\rm pBe )}
\def\totpc{\sigma_{\rm tot}(\rm pC )}
\def\totnbe{\sigma_{\rm tot}(\rm nBe )}
\def\totnc{\sigma_{\rm tot}(\rm nC )}
\def\totpibe{\sigma_{\rm tot}(\pi^{\pm} \rm Be )}
\def\totpic{\sigma_{\rm tot}(\pi^{\pm} \rm C )}
\def\tothn{\sigma_{\rm tot}( \rm hN)}
\def\tothp{\sigma_{\rm tot}( \rm hp)}
\def\totpn{\sigma_{\rm tot}( \rm pN)}
\def\totha{\sigma_{\rm tot}(\rm hA)}
\def\totpa{\sigma_{\rm tot}(\rm pA)}
\def\modnbe{\sigma_{\rm tot}^{\rm mod}(\rm Be,\sigma_{\rm tot}(\rm pp))}
\def\modnc{\sigma_{\rm tot}^{\rm mod}(\rm C,\sigma_{\rm tot}(\rm pp))}.
\def\modpimbe{\sigma_{\rm tot}^{\rm mod}(\rm Be,\sigma_{\rm tot}(\pi^- \rm p))}
\def\modpimc{\sigma_{\rm tot}^{\rm mod}(\rm C,\sigma_{\rm tot}(\pi^- \rm p))}
\def\modha{\sigma_{\rm tot}^{\rm mod}(A,\sigma_{\rm tot}(\rm hN))}
\def\modpa{\sigma_{\rm tot}^{\rm mod}(A,\sigma_{\rm tot}(\rm pp))}
\def\plab{\rm p_{\rm lab}}
\def\errextr{\delta^{\rm extr} }
\def\errbtrd{\delta^{\rm BTRD} }
\def\errfluc{\delta^{\rm fluc} } 
\def\errrate{\delta^{\rm rate} }
\def\errcont{\delta^{\rm cont} }
\def\errtgt{\delta^{\rm tgt}   }
\def\ratecorr{\Delta_{\rm T0}}
\def\contcorr{\Delta_{\rm cont}}
\begin{frontmatter}
\title{Total Cross Section Measurements  \\
       with $\pi^-$, $\Sigma^-$ and Protons \\
       on Nuclei and Nucleons around $600\, \mbox{GeV/c}$}
%
%
The SELEX Collaboration
\author[MPI]{U.~Dersch},
\author[Iowa]{N.~Akchurin},
\author[PNPI]{V.A.~Andreev},
\author[PNPI]{A.G.~Atamantchouk},
\author[Iowa]{M.~Aykac},
\author[ITEP]{M.Y.~Balatz},
\author[PNPI]{N.F.~Bondar},
\author[Trieste]{A.~Bravar},
\author[Fermi]{P.S.~Cooper},
\author[Flint]{L.J.~Dauwe},
% ...
\address[Fermi]{Fermilab, Batavia, IL 60510, U.S.A.}
% ...
\address[ITEP]{Institute of Theoretical and Experimental Physics, Moscow, Russia}
\address[MPI]{Max-Planck-Institut f\"ur Kernphysik, 69117 Heidelberg, Germany}
% ...
\address[PNPI]{Petersburg Nuclear Physics Institute, St. Petersburg, Russia}
% ...
\address[Iowa]{University of Iowa, Iowa City, IA 52242, U.S.A.}
\address[Flint]{University of Michigan-Flint, Flint, MI 48502, U.S.A.}
% ...
\address[Trieste]{University of Trieste and INFN, Trieste, Italy}
\thanks[tra]{Now at Imperial College, London SW7 2BZ, U.K.}
\thanks[trb]{Current Address: Instituto de F\'{\i}sica da Universidade Estadual de Campinas, 
\hspace*{0.4cm}UNICAMP, SP, Brazil}
\thanks[trc]{deceased}
\thanks[trd]{Present address: Dept. of Physics, Wayne State University, Detroit, MI 48201}
\thanks[tre]{Now at Universit\"at Freiburg, 79104 Freiburg, Germany}
\thanks[trf]{Now at Physik-Department, Technische Universit\"at M\"unchen, 85748 Garching, 
\hspace*{0.4cm}Germany}
\thanks[trg]{Current Address: Instituto de F\'{\i}sica Te\'{o}rica da Universidade Estadual Paulista,
\hspace*{0.4cm}S\~ao Paulo, Brazil}
\thanks[trh]{Present address: Lucent Technologies, Naperville, IL}
\thanks[tri]{Now at Max-Planck-Institut f\"ur Physik, M\"unchen, Germany}
\thanks[trj]{Present address: Motorola Inc., Schaumburg, IL}
\thanks[trk]{Present address: Deutsche Bank AG, 65760 Eschborn, Germany}
%
%
%
%
%
%
\begin{abstract}
Total cross sections for $\sim$ and $\pim$ on beryllium, carbon, 
polyethylene and copper as well as total cross sections for protons on
beryllium and carbon have been measured in a broad momentum range around
$600\, \mbox{GeV/c}$. These measurements were performed with a transmission
technique adapted to the SELEX hyperon-beam experiment at Fermilab. We report 
on results obtained for hadron-nucleus cross sections and on results for 
$\sigma_{\rm tot}(\Sigma^- \rm N)$ and $\sigma_{\rm tot}(\pi^- \rm N)$, which 
were deduced from nuclear cross sections.
\vspace*{-0.1cm}
\end{abstract}
\end{frontmatter}
%
%
%
%
\section{Introduction}
Hadronic total cross sections provide one measure of the strength of
the hadronic interaction. They have been measured for 
a variety of reactions  over a broad range of center of mass energies. 
These studies revealed that with increasing center of mass (CM) energy,
hadron-hadron cross sections (generally) decrease to a minimum and 
then start rising again. An important current physics question is
whether the rise of a specific hadron-hadron cross section is described
by a power law in the CM energy. Addressing this question requires
total cross-section experiments performed with a variety of hadronic
projectiles, targets and energies covering the maximum possible range.
However, for almost 20 years, there have been few new experiments in 
this field. Thus, important hadron-hadron cross sections as
$\sigma_{\rm tot}(\pi \rm p)$ and $\sigma_{\rm tot}(\rm Kp)$ are
measured only up to $380\, \mbox{GeV/c}$ and the total cross section
$\sigma_{\rm tot}(\Sigma^- \rm p)$ is only measured up to
$137\, \mbox{GeV/c}$. At these maximum laboratory momenta
only a first indication of the rise of these total cross
sections is observed.\\ \indent
SELEX (Fermilab E781) is a fixed-target experiment at the Fermi 
National Accelerator 
Laboratory using a hyperon beam of about $600\, \mbox{GeV/c}$. The 
SELEX spectrometer, designed for spectroscopy of charm baryons, is 
well-suited to measure total cross sections with a transmission 
technique. It has excellent scattering-angle resolution, achieved 
by a system of silicon microstrip 
detectors.\\ \indent
SELEX does not have a liquid hydrogen target. Therefore, we
measured the total hadron-nucleus cross sections $\totpimbe$, 
$\totpimc$, $\totpimpol$, 
\linebreak
$\totsimbe$, $\totsimc$, $\totsimpol$, $\totpbe$ 
and $\totpc$ with high precision. We then deduced the total cross
sections $\totsimp$ and $\totpimp$ using both a 
CH$_2$~--~C~subtraction technique and a method based on the Glauber
model to derive hadron-nucleon cross sections from hadron-nucleus 
cross sections.\\ \indent
Further, as data on hadron-nucleus cross sections are extremely 
scarce for charged projectiles, we also measured $\totpimcu$ and 
$\totsimcu$. All measurements were done during dedicated run periods 
in July 1997. Laboratory momenta range from $455\, \mbox{GeV/c}$ 
to $635\, \mbox{GeV/c}$, the highest energy yet used for these 
studies.
%
%
%
%
%
\section{Experimental setup}\label{e781_setup}
\subsection{The hyperon beam}
The hyperon beam is generated by selecting positively or negatively
charged secondaries around $600\, \mbox{GeV/c}$ that emerge from
interactions of an $800\, \mbox{GeV/c}$ primary proton beam with a
beryllium production target. Its composition has not been completely 
measured. However, we have measured the main particle components of 
the event samples, which we selected to determine total cross sections
(see section~\ref{frac_determ}). This analysis shows that at the 
position of the total cross-section target the negative beam samples 
consist in average of (52.5 $\pm$ 1.6)\% mesons and 
(47.5 $\pm$ 1.6)\% baryons. Further, we measured a $\Xi^-$ fraction
of (1.18 $\pm$ 0.06)\% in these samples. Other baryonic fractions
($\overline{\rm p }, \Omega^-$) were not measured, but empirical
formulae (see~\cite{lang95}) predict they are less than 0.1\%.
Likewise, the K$^-$ fraction of the negative beam is estimated 
with~\cite{lang95} to be (1.6 $\pm$ 1.0)\%. Thus, we expect that 
the $\pi^-$ fraction of the event samples is 
(50.9 $\pm$ 1.9)\%.\\ \indent
In the event samples for positive beam we measured a meson fraction 
of
\linebreak
(8.1 $\pm$ 1.4)\% and a baryon fraction of (91.9 $\pm$ 1.4)\%.
Furthermore, we measured a $\Sigma^+$ fraction of (2.7 $\pm$ 0.7)\%. 
Using the empirical formula given in~\cite{lang95}, we expect that 
the tiny meson fraction consists of 70\% $\pi^+$ and 30\% 
K$^+$.\\ \indent
From these compositions, one sees that as long as one can distinguish 
mesons from baryons (see section~\ref{used_section}), the SELEX hyperon 
beam offers a unique possibility to measure total cross sections for 
protons, $\pi^-$, and $\Sigma^-$ in a low contaminant environment.
\subsection{The section of the SELEX spectrometer used for 
total cross-section measurements}\label{used_section}
\begin{figure}[htbp]
  \begin{center}
  \includegraphics{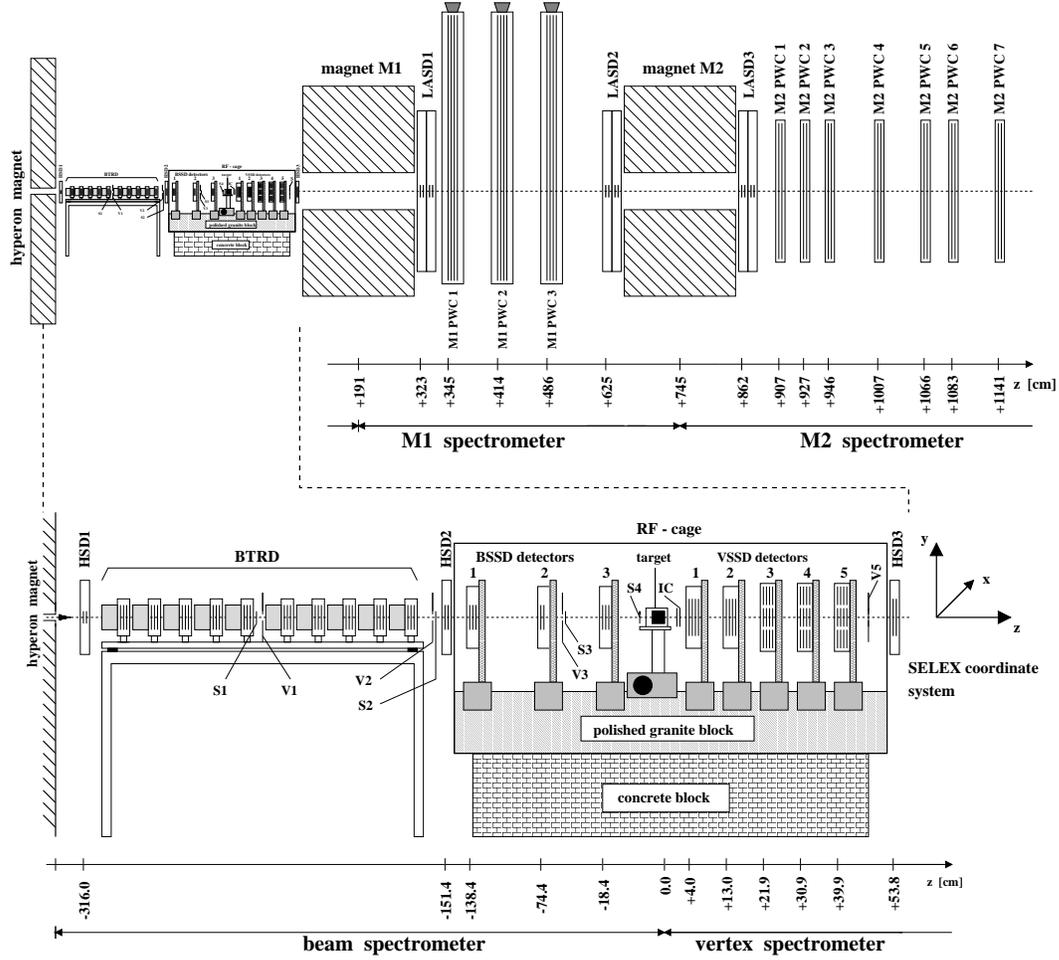}
  \vskip13cm
  \caption{Sections of the SELEX spectrometer involved in the measurement
           of total cross sections.}
  \label{fig:xsec_fullspec}
  \end{center}
\end{figure}

The SELEX spectrometer is a $60\, \mbox{m}$ long, 3~stage spectrometer. 
In total cross-section measurements, only its upstream detectors, shown 
in figure~\ref{fig:xsec_fullspec}, are used.\\ \indent
The beam spectrometer placed in front of the target, is equipped with 
12~silicon microstrip detectors to track incoming particles. The first
\noindent
4~microstrip detectors (HSDs) have a resolution (pitch/$\sqrt{12}$)
of $14.4\, \mbox{$\mu$m}$ and a maximum signal integration time of 
$100\, \mbox{ns}$. As this is the shortest integration time, but poorest
spatial resolution, of all SELEX silicon microstrip detectors, the HSDs 
serve chiefly to reject stale tracks. Always, two HSDs are housed in a 
single station. The average efficiency of the HSDs is 92\%.\\ \indent
The remaining 8~silicon microstrip detectors of the beam
spectrometer are grouped into 3~stations (BSSDs) mounted on a 
granite block inside a noise shielded cage~(RF-cage). These 
detectors have a resolution of $5.8\, \mbox{$\mu$m}$ and an 
average efficiency of 99.6\%.\\ \indent
Incoming particles are identified by a transition radiation 
detector~(BTRD) with 10~separate transition radiation detector 
modules~(TRMs). Each module is build of a radiator in succession
\begin{figure}[htbp]
  \begin{center}
  \leavevmode
  \epsfxsize=\hsize
  \epsfbox{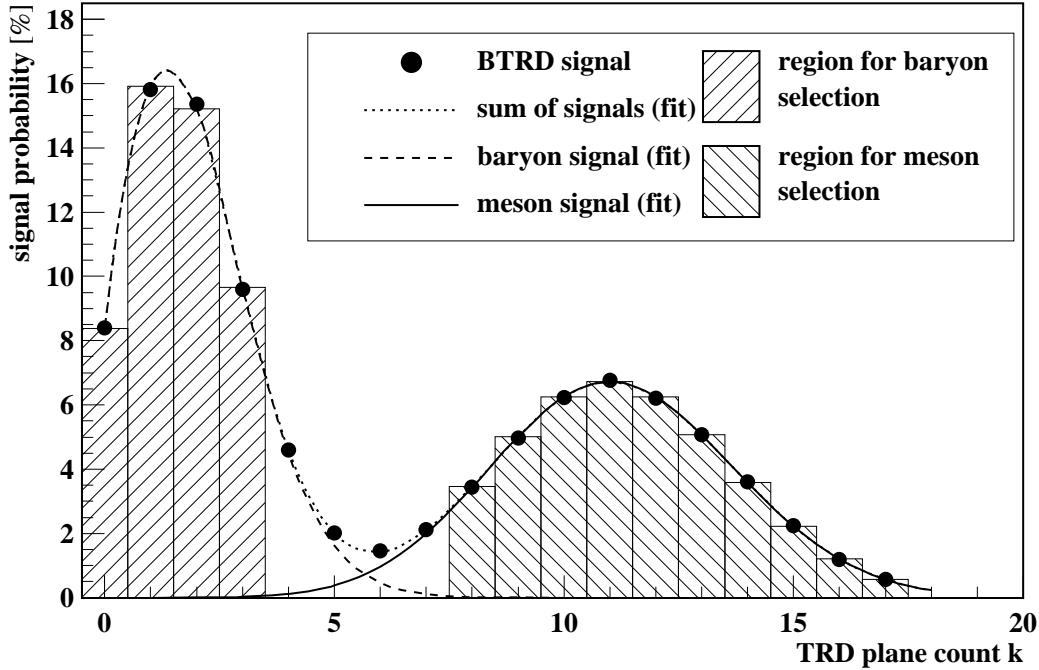}
  \end{center}
  \caption{A typical BTRD signal spectrum obtained for $600\, \mbox{GeV/c}$ 
           negatively charged secondaries.}
  \label{fig:slice}
\end{figure}
\noindent
with 3~proportional chambers~(PCs) whose operating gas is a 70\%~Xe,
30\%~CO$_2$ mixture to optimize signal response time and to
maximize absorption of transition-radiation photons.  Each 
chamber has a single anode readout amplifier.  A radiator consists 
of a stack of 200~polypropylene foils, each $17\, \mbox{$\mu$m}$ 
thick and spaced at~$500\, \mbox{$\mu$m}$.\\ \indent
Each BTRD PC gives a digital output when it detects an energy 
deposition above a fixed threshold. The sum of all PCs detecting
a signal above threshold is the TRD plane count k. A typical 
probability spectrum of TRD plane counts, a BTRD signal
spectrum, is shown in figure~\ref{fig:slice}. It shows the baryon 
and meson responses at low and high TRD plane counts, 
respectively.\\ \indent
The signal components are separated by fitting the function:
\begin{equation}
  \hspace*{-0.5cm}
  \label{btrdfit}
  {\rm p_{\rm fit}(k)} \ = \ 
  \underbrace{\sum\limits_{\rm i = 1}^{\rm 2} \kappa_{\rm i} {\rm n \choose \rm k} 
  \rm p_{\rm i}^{\rm k} \rm (1-p_{\rm i})^{\rm n-k}}_{\textnormal{baryon signal}} \ + \
  \underbrace{\sum\limits_{\rm i = 3}^{\rm 4} \kappa_{\rm i} {\rm n \choose \rm k}
  \rm p_{\rm i}^{\rm k} \rm (1-p_{\rm i})^{\rm n-k}}_{\textnormal{meson signal}}  
\end{equation}
to the normalized BTRD signal spectrum. Here, p$_{\rm i}$ and
$\kappa_i$ are fit-parameters with the constraints 
$1 = \kappa_1 + \kappa_2 + \kappa_3 + \kappa_4$ and p$_{1}$, 
p$_{2}$~$<$~p$_{3}$, p$_{4}$ and $n$ is the maximum possible
TRD plane count. The fit-pa\-ra\-me\-ters p$_{\rm i}$ 
have the meaning of a PC response probability, when a meson
(light particle) or baryon (heavy particle) 
passes. Thus, we obtain from~(\ref{btrdfit}) the meson fraction 
($\kappa_3 + \kappa_4$) and the baryon fraction ($\kappa_1 + 
\kappa_2$) of the beam.\\ \indent
The target is followed by the vertex spectrometer, which consists of 
22~silicon microstrip detectors grouped into 6~stations (VSSD1, ... , 
VSSD5 and HSD3). All VSSDs have a resolution of $5.8\, \mbox{$\mu$m}$,
and except for one plane, which has a reduced efficiency of 68\%, all 
others have an average efficiency of 98.8\%. At the end of the vertex 
spectrometer, station~HSD3 is moun\-ted to the RF~cage.\\ \indent
Although the total cross-section measurements presented in this
article are based only on detectors placed in the beam and the vertex
spectrometer, we also use other parts of the SELEX apparatus
to compute corrections. Further detectors involved in the analysis are 
situated in the M1 and the M2~spectrometer (see figure~\ref{fig:xsec_fullspec}), 
which we describe briefly:\\ \indent
The M1~spectrometer starts at the center of the M1~magnet and ends 
at the center of the M2~magnet. For high resolution tracking of high 
energy particles in the central beam region, sets of 6~silicon microstrip 
detectors (LASD1 and LASD2) are mounted to the faces of the M1 and the 
M2~magnet. The LASD detectors have a resolution of $14.4\, \mbox{$\mu$m}$, 
and an average efficiency of 95.8\%. For tracking outside the central
beam region, 12~planes of wire chambers (PWCs) are installed.\\ \indent
The M2~spectrometer starts at the center of the M2~magnet. To enhance 
the momentum resolution for high energy particles, a third station of 
silicon microstrip detectors (LASD3) is mounted to the end face of the
M2~magnet. This station is followed by 14~PWCs that are grouped into 
7~stations (M2 PWC1, ..., M2 PWC7).
\subsection{The targets}\label{targets}
To optimize the precision, total cross-section measurements are done 
with special targets. Great care was taken in selecting and machining 
adequate target materials in order to obtain best chemical and
mechanical properties (see table~\ref{tab:targets}). All targets are 
thin such that multiple scattering, quantified by 
$\sigma_{\theta}$ of Moli$\grave{\textnormal{e}}$res' formula is 
significantly lower than the 25$\, \mbox{$\mu$rad}$ angular resolution
provided by the beam and vertex spectrometer.\\ \indent
The carbon target is a stack of three quadratic pyrocarbon plates, 
each about 5~mm thick. Pyrocarbon is composed of thin carbon layers 
accumulated on top of each other in a high-temperature methane 
atmosphere. Compared to standard graphite it offers the advantages: 
no open porosity, a density close to that of a graphite monocrystal 
and less than $1\, \mbox{ppm}$ (parts per million) non-carbon 
constituents. The beam faces of the carbon plates were milled with 
a diamond-powder liquid and oriented such that the beam faces of the 
stack are parallel to each other.\\ \indent
The polyethylene target is build from a high-purity polyethylene 
granulate with less than $1000\, \mbox{ppm}$ contaminants. Molten
granulate was solidified in a vessel, where great care was taken 
that no air bubbles penetrated. The material was then carefully 
machined to a target block, and beam faces were flattened 
using a diamond pin.\\ \indent
For the beryllium and the copper target, standard industry
products of high purity are used.\\
\par
\renewcommand{\arraystretch}{1.4}
\begin{table}[htbp]
  \begin{center}
    \begin{tabular}{|c|c|cc|c|c|c|c|}
    \hline
    target               & thickness            & \multicolumn{2}{|c|}{transverse}   &  
    density              & $\sigma_{\theta}$    &  X$_{\rm coll}$      \\
    material             & $L$ [mm]             & \multicolumn{2}{|c|}{dimensions}   & 
    $\rho^*$             & [$\mu$\rm rad]           & [\%]       \\
                         & z-direction          & $x\, \mbox{[mm]}$ & $y\, \mbox{[mm]}$ 
    & [$\frac{\rm g}{\rm cm^3}$] &  &  \\[0.1cm]
    \hline
    beryllium     & 50.92   & 30.7 & 51.2  & 1.848 $\pm$ 0.002   & 8.3  & 16.86 \\
    \hline
    carbon        & 15.46   & 30.0 & 30.0  & 2.199  $\pm$ 0.003  & 6.0  & 5.40  \\
    \hline
    polyethylene  & 40.86   & 30.0 & 25.0  & 0.9291 $\pm$ 0.0008 & 6.3  & 6.66  \\
    \hline
    copper        & 1.00    & 30.0 & 30.0  & 8.96   $\pm$ 0.009  & 5.7  & 1.05  \\
    \hline
    \end{tabular}
    \vspace*{0.8cm}
    \caption{Specifics of the targets used in total
             cross-section measurements. $L$: target thickness, $\rho^*$: density,
             $\sigma_{\theta}$: expected spread in scattering angle due to 
             multiple scattering calculated with Moli$\grave{\textnormal{e}}$res' 
             formula for $\plab$~=~$600 \, \mbox{GeV/c}$, $X_{\rm coll}$: collision 
             length.}
    \label{tab:targets}
  \end{center}
\end{table}
\renewcommand{\arraystretch}{1.0}
\noindent
\subsection{Trigger and data acquisition}
The SELEX trigger is a programmable four-stage trigger, designed
to select events involving decays of charm hadrons in a 
high-intensity beam environment. The first 3~levels: T0, T1 and T2 
are hardware triggers, whereas level~T3 is an online software 
filter. In this section, we describe only the trigger as 
programmed for total cross-section data-taking.\\ \indent
At data-taking, the trigger accepted all beam 
events defined by the mi\-ni\-mum-bias condition:
\begin{equation}
  \label{trigger_def}
  T0 \ = \ S1 \wedge  S2 \wedge S3  \wedge  \overline{V1} 
  \wedge  \overline{V2}  \wedge  \overline{V3} \ .
\end{equation}
S1, S2 and S3 are scintillation counters, and V1, 
V2 and V3 are veto counters to reject beam halo (see
figure~\ref{fig:xsec_fullspec}). In definition (\ref{trigger_def}), a 
T0-pulse indicates a particle traversing the beam spectrometer in the
direction of the target beam face. Thereby, the transverse trigger acceptance 
is constrained to the size of the rectangular hole in V2 ($12.8\, \mbox{mm}\, 
\times\, 12.8\, \mbox{mm}$).\\ \indent
In order to keep the minimum bias condition provided by the definition
of T0, no information of detectors placed downstream of the experiment 
target influenced the spectrometer readout. Thus, each T0-pulse passed 
the T1 trigger level unbiased, and generated a T2-pulse, which started 
the spectrometer readout. The online software filter (level T3) was not 
used for total cross-section data-taking. Pulses of all trigger 
levels were counted by scalers for each spill, and saved in a trigger 
log file.\\ \indent
The SELEX trigger controlled readout and reset of the silicon-detector 
system, the basic tool in our total cross-section measurements. Except for 
the HSDs, all other silicon detectors use an SVX-I chip technology for data 
readout \cite{kle88}. SVX chips are controlled by a sequencer SRS (silicon 
readout sequencer) that interacts very closely with the trigger. First,
it keeps the silicon detectors sensitive (for about $5\, \mbox{$\mu s$}$ 
effective integration time) and starts the chip readout when receiving a
T2-pulse. Second, the SRS resets the SVX-chips, when a silicon-clear 
signal arrives. Thereby, the silicon-clear signal is generated in the 
trigger logic through:
\begin{equation}
  \label{reset_def}
  \textnormal{Silicon clear} \ = V5_{\rm mult} \ \vee \ C_{\rm pulser} \ 
  \vee \ (T1 \ \wedge \ \overline{T2}) \ .
\end{equation}
Here, $C_{\rm pulser}$ are pulses from a gate generator running at a
frequency of $20\, \mbox{kHz}$ and $V5_{\rm mult}$ represents pulses 
generated, when the V5 veto counter (see figure~\ref{fig:xsec_fullspec})
detects a high multiplicity event.
The condition~$(T1 \wedge \overline{T2})$ was irrelevant for total 
cross-section data.
\subsection{Experimental conditions and recorded data}
During the fixed-target run 1996/97, the TEVATRON was operated in 
$60\, \mbox{s}$ cycles with a spill time of $20\, \mbox{s}$. Data 
for total cross sections were taken during dedicated periods, with
optimized experimental conditions for this 
measurement.\\ \indent
By adjusting the flux of the $800\, \mbox{GeV/c}$ proton beam, 
the T0-rate was optimized to run the SELEX DAQ near, but safely 
below its capacity limit of 5$\cdot$10$^{4}$ particles per 
spill. The low hyperon-beam flux allowed a high silicon-clear rate, 
which resulted in a very low-noise condition for the silicon-detector 
system and a low probability for stale tracks.\\ \indent 
During data-taking, the M1~magnet was switched off to obtain a
$2.5\, \mbox{m}$ field- and material-free section, serving as 
fiducial region for precise reconstruction of hyperon decays. 
Magnet~M2 was operated with a transverse momentum kick of 
$\rm p_{\rm T}^{\rm M2} = 0.84\, \mbox{GeV/c}$.\\ \indent
At data-taking start, after mounting an experiment target in the 
RF-cage, an alignment RUN was taken to account for eventual 
detector displacements caused during the target installation. Then, 
the position of the experiment target was alternated every
$30\, \mbox{min}$ between its out and in-beam position. Thus, almost 
equal amounts of data were taken with full and empty target. A RUN, 
started after each target-position change, comprised typically 
10$^6$ events. A total of 9.8$\cdot$10$^7$ minimum-bias events 
were recorded with negative beam for the targets Be, C, Cu and CH$_2$.
With positive beam, 3.0$\cdot$10$^7$ minimum-bias events were written 
using the targets Be and C.
%
%
%
%
\section{The principle of the transmission method}\label{principle}
In contrast to scattering experiments, where $\sigma_{\rm tot}$ is deduced
from a measured scattering angle distribution, in a transmission experiment 
$\sigma_{\rm tot}$ is deduced from the number of unscattered projectiles. 
Strictly, unscattered means zero scattering angle, but experimental resolution 
and Coulomb scattering limit this to a determination of the number of projectiles
scattered by an angle $\theta$, which is smaller than a maximum angle parameter 
$\theta_{\rm max}$ (F$_{\rm o}(<\theta_{\rm max})$). Thus, one infers 
the number of unscattered particles by extrapolating 
F$_{\rm o}(< \theta_{\rm max})$ to $\theta_{\rm max}$~=~0.\\ \indent
A standard transmission experiment consists of three elements: 
beam monitor, target, and transmission counter. The number of projectiles 
hitting the target under full-target (empty-target) condition F$_{\rm o}$ 
(E$_{\rm o}$) is counted by the beam monitor placed in front of the target. 
A transmission counter, placed downstream of the target, counts the corresponding 
number of projectiles F$_{\rm tr}(<\Omega_{\rm i})$ (E$_{\rm tr}(<\Omega_{\rm i})$), 
leaving the target within the maximum solid angles $\Omega_{1}$ ... 
$\Omega_{\rm N}$. Recorded counts are combined to a set of partial cross sections 
$\sigma_{\rm part}(<\Omega_{\rm i})$, defined as:
\begin{equation}
\label{xpart1}
\! \! \! \! 
\sigma_{\rm part}(<\Omega_{\rm i}) \ = \ \frac{1}{\rho L} \log \left[
\frac{\rm F_{\rm o}}{\rm F_{\rm tr}(<\Omega_{\rm i})} 
\frac{\rm E_{\rm tr}(<\Omega_{\rm i})}{\rm E_{\rm o}} \right] \ \ \ \
\ \  \textnormal{with} \ \ \
\rho = \frac{\rm N_{\rm A} \rho^*}{A} \ ,
\end{equation}
where $\rho$ is the density of scattering centers in the target, $A$ is the
atomic mass and $\rm N_{\rm A}$ is Avogadro's number.\\ \indent
Driving our choice of a transmission method is an important technical 
advantage of equation~(\ref{xpart1}).
We do not need to know absolute efficiencies of the beam and the transmission 
monitor. Their absolute values will cancel in~(\ref{xpart1}) as long as they 
remain unchanged between and during the full- and the empty-target RUNs 
(stability condition).\\ \indent
Taking into account the event correlations between F$_{\rm o}$ 
(E$_{\rm o}$) and F$_{\rm tr}(<\Omega_{\rm i})$ (E$_{\rm tr}(<\Omega_{\rm i})$), 
the statistical error of a partial cross section is given by:
\begin{equation}
\label{error_part}
\delta \sigma_{\rm part}(<\Omega_{\rm i}) \ = \ \frac{1}{\rho L} \sqrt{ 
\frac{1}{\rm F_{\rm tr}(<\Omega_{\rm i})} - \frac{1}{\rm F_{\rm o}} +
\frac{1}{\rm E_{\rm tr}(<\Omega_{\rm i})} - \frac{1}{\rm E_{\rm o}} } \ .
\end{equation}
In a thin target approximation ($\rho L \sigma_{\rm tot} \ll 1$), a partial
cross section $\sigma_{\rm part}(<\Omega_{\rm i})$ is related to the total
hadronic cross section $\sigma_{\rm tot}$ (see e.g.~\cite{der98}) by:
\begin{eqnarray}
  \label{relation}
  \nonumber
  \sigma_{\rm tot} & = & \sigma_{\rm part}(<\Omega_{\rm i}) \ \underbrace{ - \
  \int_{\Omega_{\rm i}}^{4\pi} \left( \frac{d\sigma}{d\Omega} \right)_{\rm C} 
  d\Omega \  - \
  \int_{\Omega_{\rm i}}^{4\pi} \left( \frac{d\sigma}{d\Omega} \right)_{\rm CN} 
  d\Omega}_{\textnormal{Correction for C~and CN~scattering}} \\
              & & \hspace{2.2cm} + \ \underbrace{
  \int_{0}^{\Omega_{\rm i}} \left( \frac{d\sigma}{d\Omega} \right)_{\rm el}^{\rm hadr} 
  d\Omega}_{\textnormal{elastic term}} \ + \
  \underbrace{
  \int_{0}^{\Omega_{\rm i}} 
  \left( \frac{d\sigma}{d\Omega} \right)_{\rm inel}^{\rm hadr} 
  d\Omega}_{\textnormal{inelastic term}} \ .
\end{eqnarray}
In equation (\ref{relation}), $\sigma_{\rm tot}$ is infered by 
first correcting partial cross sections for Coulomb~scat\-te\-ring (C) and 
the Coulomb hadronic interference (CN) and then extrapolating to zero solid 
angle.
%
%
%
%
\section{Data analysis}\label{data_analysis}
\subsection{Data selection}\label{data_selection}
In general, total cross-section data taken for a specific target
were subject to varying experimental conditions: thresholds
on silicon microstrip detectors, high voltages for trigger scintillators,
and the inclination angle between primary proton beam and production target. 
Therefore, data belonging to a cross-section measurement with a specific 
target were divided into as many data sets as differing 
conditions had to be taken into account. This offered the possibility 
to calculate corrections and errors specifically for each experimental 
condition in a later stage of the analysis. To preserve the stability 
condition mentioned in chapter~\ref{principle}, a spill by spill 
data pre-selection was performed. Data of a spill or a whole run were 
rejected:
\begin{enumerate}
\item When the experimental conditions concerning the functionality
      of the spectrometer (detector efficiencies, trigger
      performance and track reconstruction efficiencies) suddenly
      changed.
\item When it was not possible to synchronize raw data with
      information in the trigger log file.
\item When the BTRD showed instabilities or when the beam phase
      space lay outside the BTRD fiducial region.
\end{enumerate}
\subsection{Event selection for normalization}\label{norm}
The total cross-section determination is made by counting how many
good beam tracks are removed from the beam by interactions in 
the target. The normalization therefore depends only on the
number of good beam tracks, which are identified by a software 
decision routine. This routine reconstructs tracks in the beam 
spectrometer using the HSD and BSSD hit information. It preserves 
the minimum-bias condition for the selected data by strictly 
avoiding event-selection rules that require information from
detectors placed downstream of the target. An event is accepted 
when it is possible to reconstruct a track called a ``norm 
track'', provided that the following properties are satisfied:
\begin{enumerate}
\item Not more than a total of 150~hits in all BSSDs.
\item At least 6~hits from BSSD planes along the track.
\item At least one hit from an HSD plane along the track
      (HSD-tagging).
\item A reduced track-fit $\chi^2$ below~3.
\item An extrapolated origin of the track at the known
      transverse position of the primary production target.
\item Track intercept and slope parameters within the
      beam phase space accepted by magnetic collimation.
\item A transverse track position at the longitudinal 
      position of the experimental target, which is inside 
      the trigger acceptance window and inside BTRD
      acceptance.
\item A beam momentum assigned to the track, which is $\pm$ 
      $100\, \mbox{GeV/c}$ around the center of 
      gravity value of the momentum spectrum.
\end{enumerate}
Condition~{3} rejects stale tracks. The selection 
rules~{4~--~6} remove events in which hyperons decay before 
reaching the experiment target or react with detector material 
in the beam spectrometer. Constraint~{7} assures also that 
selected tracks point to the mid-part of the experiment 
target face, where the best mechanical accuracy is 
obtained.\\ \indent
All listed conditions were true for about~50\% of the selected 
events. From the resulting set of norm tracks for full- and empty-target 
conditions, we establish classes of BTRD-tagged norm tracks. 
This is done by introducing cuts on the BTRD information as indicated 
in figure~\ref{fig:slice} to separate baryonic and mesonic
norm tracks. We then determine the corresponding normalization 
counts F$_{\rm o}$ and E$_{\rm o}$ by summing the norm tracks over 
the appropriate signal region.
\subsection{Transmission counting}
When a norm track is found in the event, we reconstruct a single
track in the vertex spectrometer, which is leaving the interaction 
target at small angle with respect to the norm track.\\ \indent
The single-track algorithm was efficient and fast. It used
hits of HSD3 to remove stale tracks. With loose cuts on the 
track parameters 98\% of the norm tracks got assigned a track 
in the vertex spectrometer. Such vertex tracks were finally 
accepted as ``transmitted tracks'', when:
\begin{enumerate}
\item There are at least 15 hits from VSSDs found within a 
      track search corridor.
\item The reduced track-fit $\chi^2$ is below~3.
\end{enumerate}
For each transmitted track, the scattering angle~$\theta$ between norm and 
transmitted track is calculated. Following the idea of~\cite{bia81}, a 
four-momentum transfer $t$ is assigned to the event using the small angle 
approximation $t \ \approx \ - \rm p_{\rm beam}^2 \theta^2$, where 
p$_{\rm beam}$ is the momentum of the incoming particle. Transmitted tracks 
are assigned to $t$ bins of width $5.0\cdot 10^{-4}\, \mbox{GeV$^2$/c$^2$}$. 
Note that we count transmitted tracks in $t$-bins, rather than in bins
of solid angle $\Omega$ as discussed in chapter~\ref{principle}.
Summing the events in the $t$-bins from zero up to a maximum $t_{\rm i}$ 
leads to sets of transmission counts F$_{\rm tr}(<|t_{\rm i}|)$ and 
E$_{\rm tr}(<|t_{\rm i}|)$.
\subsection{Spectra of uncorrected partial cross sections}
Using the counts F$_{\rm o}$, E$_{\rm o}$, F$_{\rm tr}(<|t_{\rm i}|)$, 
E$_{\rm tr}(<|t_{\rm i}|)$ and the mechanical properties of the 
\begin{figure}[h]
  \begin{center}
  \leavevmode
  \epsfxsize=\hsize
  \epsfbox{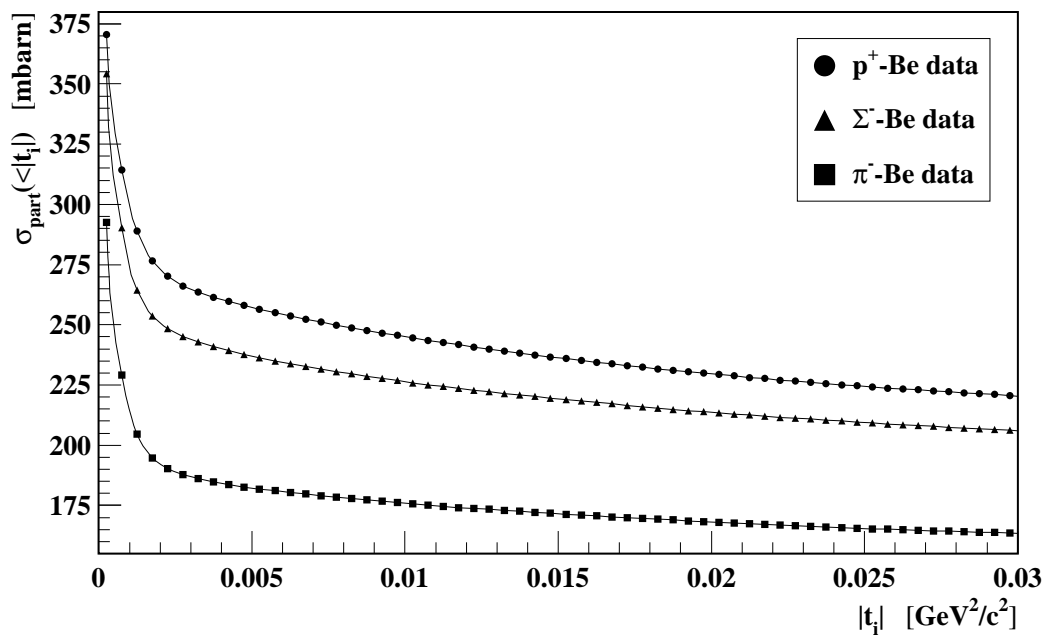}
  \end{center}
  \caption{Spectra of uncorrected partial cross sections 
           resulting from beryllium target data sets.} 
  \label{fig:xpart_spectra}
\end{figure}
targets, partial cross sections $\sigma_{\rm part}(<|t_{\rm i}|)$ are 
calculated according to equation~(\ref{xpart1}).\\ \indent
Figure \ref{fig:xpart_spectra} shows some spectra for uncorrected 
partial cross sections. The strong rise of $\sigma_{\rm part}(<|t_{\rm i}|)$ 
for $|t|$~$<$~$0.002\, \mbox{GeV$^2$/c$^2$}$ is ascribed to multiple 
scattering in the target and the finite angular resolution of 
$\approx 25\,\mbox{$\mu$rad}$. Differing levels of partial cross-section 
spectra for beam particles of different kind indicate nicely the dependence 
of the total cross section on the projectile type.
\subsection{Corrections for non-hadronic effects}
Partial cross sections were corrected for single Coulomb scattering (C)
and for the Coulomb-Nuclear interference effect (CN) evaluating the 
expression:
\begin{equation}
  \hspace*{-1cm}
  \sigma_{\rm part}^{\rm corr}(<|t_{\rm i}|) \ = \  
  \sigma_{\rm part}(<|t_{\rm i}|) \ - \! \! \!
  \underbrace{  \int_{-\infty}^{t_{\rm i}} 
  \left( \frac{d\sigma}{dt'} \right)_{\rm C} 
  dt'}_{\textnormal{C correction} } \  - \! \! \!
  \underbrace{\int_{-\infty}^{t_{\rm i}}   
  \left( \frac{d\sigma}{dt'} \right)_{\rm CN} 
  dt'}_{\textnormal{CN correction}} \ .
\end{equation}
\noindent
Applying the Coulomb correction, a change in the extrapolated cross section
of not more than 0.5\% is observed for the light targets Be, C and CH$_2$. 
For the Cu~target a change of up to 11\% is noticed. The CN~correction 
is roughly one order of magnitude smaller than the Coulomb correction and 
has negligible effect on the extrapolated cross section.
\subsection{The extrapolation method}
As $|t|$ approaches zero, the growth behavior of partial cross 
sections is governed ideally by the elastic term in 
equation~(\ref{relation}). At small $|t|$, hadronic coherent 
elastic scattering off nuclei dominates. Thus, we obtain for 
the elastic term in equation~(\ref{relation}), the expression:
\begin{equation}
   \label{intelhad}
   \int_{0}^{t} \left( \frac{d\sigma}{dt'} \right)_{\rm el}^{\rm hadr} 
   dt' \ = \ \frac{\sigma_{\rm tot}^2}{16 \pi B_{\rm nuc}} (1 \ + \ \rho^2)
  \left[1 \ - \ e^{B_{\rm nuc}t} \right] \ ,
\end{equation}
where $B_{\rm nuc}$ is the exponential slope observed in hadronic
coherent elastic scattering off nuclei. Therefore, we choose 
the functional form
\begin{equation}
   \label{schiz_param}
   f(\alpha_1, \alpha_2, t) \ = \ \alpha_1  \left[1 \ - \ e^{\alpha_2 t} 
  \right] 
\end{equation}
to describe the variation of partial cross sections with respect
to $|t_{\rm i}|$.\\ \indent
The parameters $\alpha_1$ and $\alpha_2$ are determined in fitting 
function~(\ref{schiz_param}) to differences in corrected partial cross 
sections of adjacent $t$-bins. Thereby, only those corrected partial 
cross sections where $t_{\rm i}$ is in the range of 
$t_{\rm min}$~=~-$0.007\, \mbox{GeV$^2$/c$^2$}$ to $t_{\rm max}$~=~-$0.03\, 
\mbox{GeV$^2$/c$^2$}$ enter the fit procedure. The limits $t_{\rm max}$ 
and $t_{\rm min}$ account for experimental sensitivity to hadronic
coherent elastic scattering off nuclei. Their derivation is described 
in section~\ref{tmin_tmax}.\\ \indent
Starting from the partial cross section $\sigma_{\rm part}(<|t_{\rm min}|)$,
the extrapolation to the total cross section $\sigma_{\rm tot}$ is 
determined by accounting for the expected growth in partial cross sections
from $t_{\rm min}$ to $t$~=~0 using the expression:
\begin{equation}
  \label{xextrap}
  \sigma_{\rm tot} \ = \ \sigma_{\rm part}(<|t_{\rm min}|) \ + \ \alpha_1   
                     \left[1 \ - \ e^{\alpha_2 t_{\rm min}} \right] \ .
\end{equation}
\subsubsection{The limits $t_{\rm min}$ and $t_{\rm max}$ and the 
sensitivity of the SELEX experiment to coherent hadronic
elastic scattering off nuclei}\label{tmin_tmax}
\begin{figure}[htbp]
  \begin{center}
  \leavevmode
  \epsfxsize=\hsize
  \epsfbox{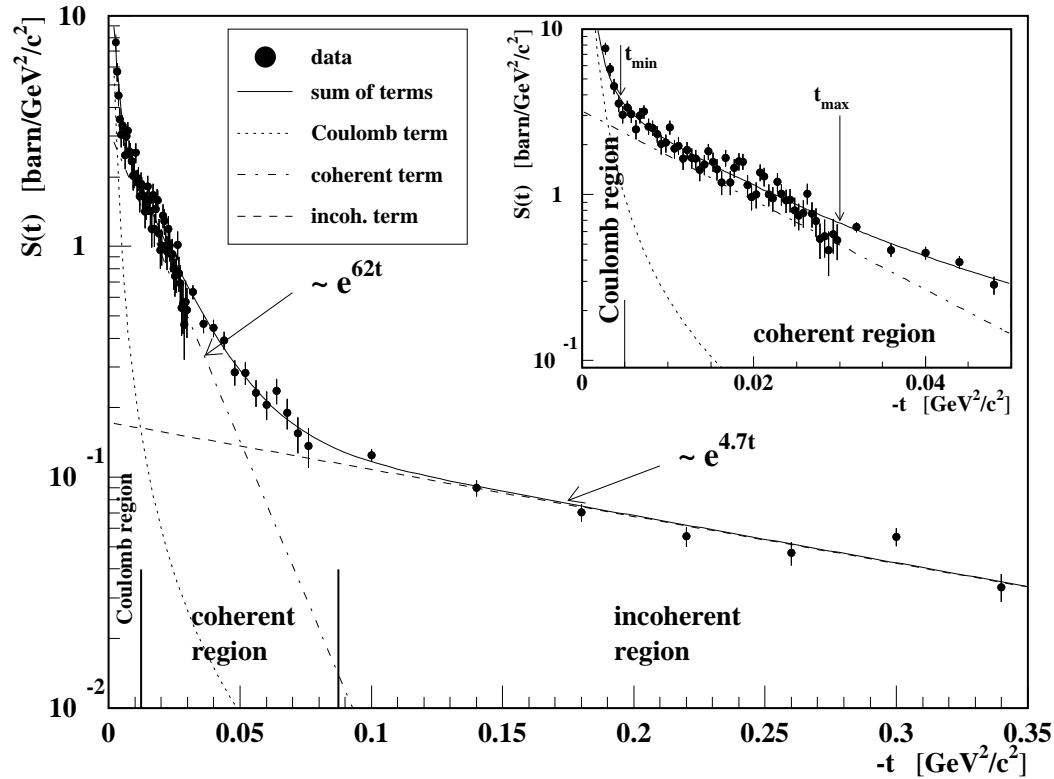}
  \end{center}
  \caption{Differential scattering spectrum obtained for $\Sigma^-$ carbon
           reactions, showing the Coulomb, the coherent and the
           incoherent region.}
  \label{fig:xdiff}
\end{figure}

In measurements of hadron-nucleus cross sections, it is essential 
that the experiment is sensitive to hadronic coherent elastic 
scattering off nuclei. Further, one must be able to distinguish 
coherent from incoherent scattering processes off nucleons. 
In scattering off nuclei, the nucleus can break up when the energy 
transfer exceeds the binding energy of its nucleons. This leads to 
a contribution of incoherent scattering off nucleons for 
$|t|$~$>$~0.015\, \mbox{GeV$^2$/c$^2$}. In that case, the hadronic 
differential elastic cross section, entering the elastic term of 
equation~(\ref{relation}), contains two parts:
\begin{equation}
  \label{quasiel}
  \! \! \! \!
  \left( \frac{d\sigma}{dt} \right)_{\rm el}^{\rm hadr} \ = \ 
  \underbrace{
  \frac{\sigma_{\rm tot}^2(\rm hA)}{16\pi} (1 \ + \ \rho'^2) 
  e^{B_{\rm nuc}t}
  }_{\textnormal{coherent scattering}} \ + \ 
  \underbrace{\rm N(A) \frac{\sigma_{\rm tot}^2(\rm hN)}{16\pi}
     e^{B_{\rm N}t}}_{\textnormal{incoherent scattering}} \ .
\end{equation}
There is a term for coherent elastic scattering off the nucleus, 
in which $\totha$ is the total nuclear cross section, and a term
for incoherent scattering off nucleons, in which $\sigma_{\rm tot}(\rm hN)$ 
is the corresponding hadron-nucleon cross section. $B_{\rm N}$ is the
slope parameter for scattering off nucleons, and $\rm N(A)$ is a factor 
describing the effective number of nucleons taking part in the incoherent 
process for target nuclei of mass $A$ 
(see~\cite{bell66}).\\ \indent
The contribution of the incoherent term decreases the growth behavior of 
the elastic term in~(\ref{relation}) because $B_{\rm N}$ is typically one 
or more orders of magnitude smaller than $B_{\rm nuc}$. An extrapolation 
based on partial cross sections, selected in a $|t|$-range far above 
$0.015\, \mbox{GeV$^2$/c$^2$}$ would lead to a systematically lowered 
cross-section result as a fraction of the elastic processes is not 
discriminated. Consequently, we looked for a $t$-interval 
[$t_{\rm min}$; $t_{\rm max}$] to select partial cross sections where 
their growth is dominated by $B_{\rm nuc}$.\\ \indent
The sensitivity of the SELEX spectrometer to hadronic coherent
elastic scattering off nuclei was verified by looking at 
background subtracted but not acceptance corrected differential 
scattering spectra, which are defined by:
\begin{equation}
  \label{elastic}
  {\rm S(t)} \ = \ \frac{1}{\rho L \Gamma} 
               \left[\frac{\rm F(t)}{\rm F_{\rm o}} \ - \
               \frac{\rm E(t)}{\rm E_{\rm o}} \right] \ \ .
\end{equation}
Here, $\Gamma$ is the width of the $t$-bins. F($t$) (E($t$)) 
is the number of scattering events found in the full-target 
(empty-target) data sets that fall into the interval 
[$|t| - \Gamma/2$ ; $|t| + \Gamma/2$].\\ \indent
Figure~\ref{fig:xdiff} shows a typical example of an S($t$)~spectrum 
obtained for $\Sigma^-$ scattering off carbon nuclei. The spectrum shows 
three regions governed by apparently different exponential slopes, which
can be explained by contributions of Coulomb scattering, coherent elastic 
scattering and incoherent elastic scattering comparable to measurements 
described in~\cite{bell66}.\\ \indent
Determinations of the slope parameters $B_{\rm nuc}$ and $B_{\rm N}$ 
in S($t$) spectra showed the expected order of magnitude for all targets,
and $B_{\rm nuc}$ agreed quite well with data presented in~\cite{sch80}.
Furthermore, the magnitude of $B_{\rm nuc}$ is also reflected by the size 
of parameter $\alpha_2$ in equation~(\ref{xextrap}), when applying the 
extrapolation.\\ \indent
From such studies, we choose $t_{\rm max}$~=~-$0.03\, 
\mbox{GeV$^2$/c$^2$}$ fixed, as this value is well inside the region 
dominated by coherent hadronic elastic 
scattering off nuclei for all targets. The contribution of the integrated 
incoherent term at this $t_{\rm max}$ is much lower than the integrated 
coherent term.\\ \indent
Further, to avoid large multiple-scattering corrections, we choose a
$t_{\rm min}$ of $-0.007$ $\mbox{GeV$^2$/c$^2$}$, such that the angular 
resolution has negligible effect on the ex\-tra\-pola\-ted total cross 
section. 
%
%
%
%
%
%
\section{Corrections}\label{data_corrections}
\subsection{Trigger-rate corrections}\label{rate_correction}\label{t0ratecorr}
\begin{figure}[htbp]
  \begin{center}
  \leavevmode
  \epsfxsize=\hsize
  \epsfbox{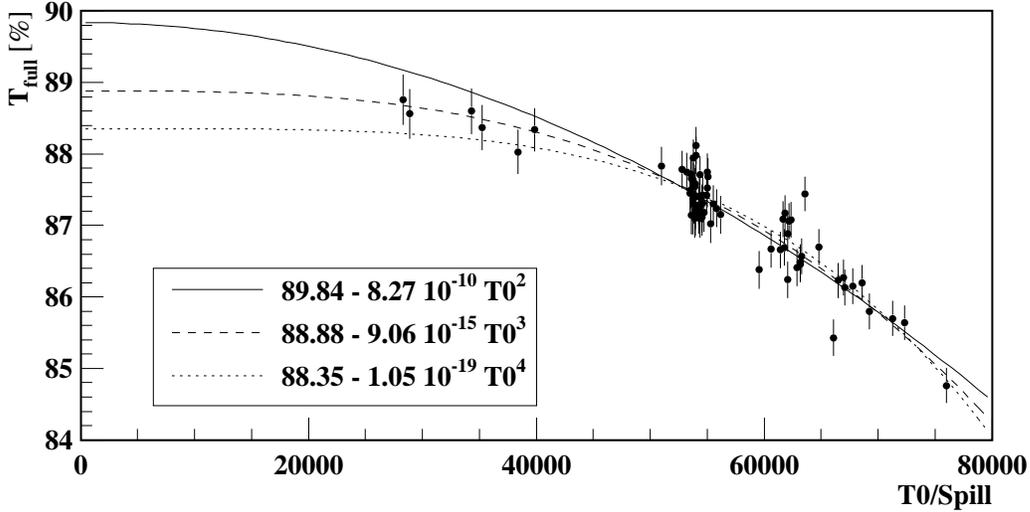}
  \end{center}
  \caption{Dependency of full-target transmission ratios on
           the T0-count.}
  \label{fig:rate_effect}
\end{figure}

The trigger rate influences the reconstruction efficiency for tracks 
and thus alters the transmission ratios $\rm T_{\rm full}$ and 
$\rm T_{\rm empty}$ per spill. Figure~\ref{fig:rate_effect} shows an 
instructive example of this effect.\\ \indent
Due to the rate effect, our extrapolated total cross-section 
experiences a shift~$\Delta_{\rm T0}$ when the average T0-counts,
calculated for all empty and all full-target spills separately, 
differ.\\
\par
\noindent
To determine the shift~$\Delta_{\rm T0}$ we calculate full and
empty-target transmission ratios per spill for $|t|$~$<$~$0.01\, 
\mbox{GeV$^2$/c$^2$}$ and describe their rate dependency by fitting 
to the expression
\begin{equation}
  \label{ratefit}
  \rm T_{\rm fit}(T0) \ = \ \tilde{\beta}_{\rm 1,k} \ + \ 
  \tilde{\beta}_{\rm 2,k} T0^{\rm k} \ .
\end{equation}
We have studied the effect of different powers (k~=~2,~3,~4)
to estimate systematic errors.\\ \indent
We choose the average T0-rate $\overline{\rm T0}$, comprising 
all full and all empty-target spills as reference rate for the 
rate correction. Thus, transmission ratios per spill are 
corrected by evaluating:
\begin{eqnarray}
  \underbrace{\ \ T_{\rm j,k}^{\rm T0} \ \ 
  }_{\textnormal{corrected}} & = & \ 
  \underbrace{T_{\rm j}(|t|<0.01 \textnormal{GeV$^2$/c$^2$} 
  )}_{\textnormal{uncorrected}} \ + \ 
  \underbrace{\tilde{\beta}_{\rm 2,k}(\overline{\rm T0}^{\rm k} - 
  {\rm T0}_{\rm j}^{\rm k})}_{\textnormal{correction}} \ ,
\end{eqnarray}
which results in a set of corrected transmission ratios 
$\rm T_{\rm j,k}^{\rm T0}$. Fit-function dependent offsets 
$\Delta_{\rm T0,k}$ are deduced by:
\begin{equation}
  \Delta_{\rm T0,k} \ = \ \sigma_{\rm part}^{\rm T0,k}(<0.01 
  \textnormal{GeV$^2$/c$^2$})
  \ - \ \sigma_{\rm part}(<0.01 \textnormal{GeV$^2$/c$^2$})
\end{equation}
and averaged to a mean offset $\Delta_{\rm T0}$. Total cross 
sections are then corrected by:
\begin{equation}
  \sigma_{\rm tot}^{\rm T0} \ = \ \sigma_{\rm tot} \ + \ 
  \Delta_{\rm T0} \ .
\end{equation}
Averaged sizes of the rate correction are presented in
table~\ref{tab:error_sizes}. We want to mention that the 
copper data were taken at higher rate, where the slope of 
function~(\ref{ratefit}) is steeper. This, together with 
the small thickness of the copper target, causes large 
corrections.
\subsection{Corrections for beam contaminants}\label{cont_corr}
A transition radiation detector does not make an exact particle 
identification because of statistical fluctuations in X-ray
generation and background from various processes. Therefore, 
when selecting the baryon or the meson component of the 
hyperon beam by applying cuts on the BTRD plane count, we 
need to account for:
\begin{enumerate}
\item The amount of meson (baryon) contaminants in the baryon
      (meson) sample and the effect on the total cross section.
\item The amount of baryon (meson) contaminants in a specific
      sample for a measurement with protons or $\Sigma^-$ ($\pi^-$) 
      and the effect on the total cross section.
\end{enumerate}
Once the contaminant fraction $\epsilon$ is determined, 
the experimental cross section $\sigma_{\rm tot}^{\rm exp}$
can be corrected by the term $\Delta_{\rm cont}$ using:
\begin{equation}
  \label{contcorr}
  \sigma_{\rm tot}^{(1)} \ = \ \sigma_{\rm tot}^{\rm exp}  
  \ + \ \underbrace{ \frac{1}{\rho L}
  \log \left[ 1 \ + \ \epsilon^{(2)} (e^{-\rho L 
  (\sigma_{\rm tot}^{(2)} \ - \ 
  \sigma_{\rm tot}^{(1)})} \ - \ 1 ) \right]}_{
  \textnormal{Correction} \ \Delta_{\rm cont}   } \ .
\end{equation}
This formula was derived in~\cite{murt75} for a two component 
beam having a contamination fraction $\epsilon^{(2)}$.
\subsubsection{Beam contaminant determination}\label{frac_determ}
\begin{figure}[htbp]
  \begin{center}
  \leavevmode
  \epsfxsize=\hsize
  \epsfbox{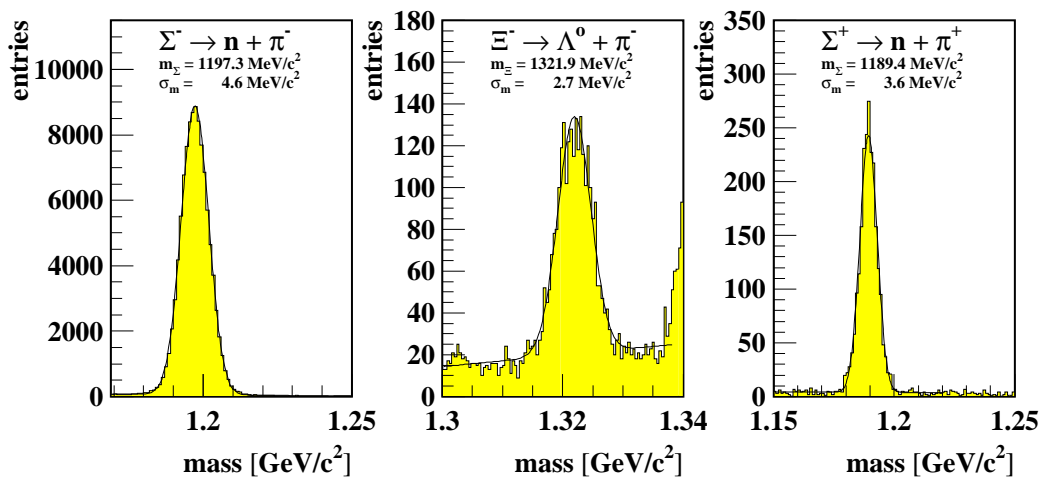}
  \end{center}
  \caption{Hyperon-mass spectra obtained from 
           reconstructed $\Sigma^-$, $\Xi^-$ and 
           $\Sigma^+$ decays. The spectra are fit to
           a Gaussian plus a linear background function. 
           m$_{\rm X}$ is the mean mass found for hyperon X and 
           $\sigma_{\rm m}$ is the corresponding mass resolution
           resulting from the fit.}
  \label{fig:mass_spectra}
\end{figure}

In a first step, total cross sections resulting from data sets
are corrected for the fraction of mesons (baryons) in a baryon
sample (meson sample) using~(\ref{contcorr}). Therefore, we
fit function~(\ref{btrdfit}) to normalized BTRD signal spectra,
which are recorded for norm tracks. For negative beam these fits 
yield an average baryon fraction ($\kappa_1$ + $\kappa_2$) of 
(47.5 $\pm$ 1.6)\% and an average meson fraction 
($\kappa_3$ + $\kappa_4$) of (52.5 $\pm$ 1.6)\%. For positive 
beam, we measure a baryon fraction of (91.9 $\pm$ 1.4)\% and a 
meson fraction of (8.1 $\pm$ 1.4)\%. To deduce the meson 
(baryon) contaminant fraction~$\epsilon$, we sum the meson 
(baryon) component of~(\ref{btrdfit}) over the TRD plane 
count region shown in figure~\ref{fig:slice}. Further, the 
difference~$\sigma_{\rm tot}^{(2)} \ - \ \sigma_{\rm tot}^{(1)}$ 
is calculated in taking rate corrected extrapolated cross-section 
results obtained for the meson and the baryon beam 
component. \\ \indent
In a second step, we account for the main contaminant disturbing a
specific measurement for protons, $\Sigma^-$ and $\pi^-$. According 
to the expected hyperon-beam composition we correct:
\begin{enumerate}
\item For the effect of $\Xi^-$ particles in the baryon sample,
      when measuring $\Sigma^-$A cross sections.
\item For the effect of $\Sigma^+$ particles in the baryon sample,
      when measuring pA cross sections.
\item For the effect of K$^-$ particles in the meson sample,
      when measuring $\pi^-$A cross sections.
\end{enumerate}
For case~(1), we measure the overall fraction of $\Xi^-$~particles 
in each negative-beam data sample and for case~(2) we measure the 
overall fraction of $\Sigma^+$~particles in each positive-beam data 
sample. Therefore, we count the decays $\Sigma^- \rightarrow$ n + 
$\pi^-$, $\Xi^- \rightarrow \Lambda^o + \pi^-$ and $\Sigma^+ 
\rightarrow$ n + $\pi^+$, reconstructed for a known amount 
of norm tracks within the field-free region of the M1~magnet. 
Figure~\ref{fig:mass_spectra} shows some hyperon-mass spectra 
obtained by the decay reconstruction.\\
\par
\noindent
Particle decay counts are 
corrected for geometrical acceptance, branching ratio and 
decay losses after the target, to yield the overall hyperon
contaminant fractions. Here, we find an overall $\Xi^-$~fraction
of (1.18 $\pm$ 0.06)\%, and an overall $\Sigma^+$~fraction of 
(2.7 $\pm$ 0.7)\%. These fractions are then divided by the baryon 
fraction ($\kappa_1$ + $\kappa_2$), known from the first step 
procedure to yield the hyperon contaminant fraction~$\epsilon$ 
of the baryon component.\\ \indent
Case~(3) requires knowledge of the number of K$^-$ particles in the meson 
sample. As the SELEX spectrometer does not provide a tool to discriminate
$600\, \mbox{GeV/c}$ $\pi^-$ against K$^-$ particles, we estimate the
overall fraction of K$^-$ particles in the sample using particle-flux
parameterizations of~\cite{lang95}. This results in an overall 
K$^-$~fraction of (1.6 $\pm$ 1.0)\%, which divided by the meson fraction 
($\kappa_3$ + $\kappa_4$) yields the K$^-$~contaminant fraction of the
meson component.\\ \indent
Calculating the contaminant correction using equation~(\ref{contcorr}) 
requires know\-ledge of the total cross sections $\sigma_{\rm tot}(\Xi^-\rm A)$,
$\sigma_{\rm tot}(\Sigma^+\rm A)$ and $\sigma_{\rm tot}(\rm K^- \rm A)$. As 
data on these cross sections are either scarce or do not exist, we estimate
them using approximations like:
\begin{equation}
\label{estimator}
\sigma_{\rm tot}(\Xi^- \rm A) \ \approx \ \sigma_{\rm tot}(\Xi^- \rm p)
\frac{\sigma_{\rm tot}(\rm pA)}{\sigma_{\rm tot}(\rm pp)} \ ,
\end{equation}
and neglect weak but existing energy dependencies. Necessary data for
hadron-nucleon cross sections are taken from~\cite{bia81,pdg98} and data 
for pA-cross sections are taken from~\cite{sch79}.\\ \indent
Averaged sizes of the contaminant correction including both correction 
steps are shown in table~\ref{tab:error_sizes}.
%
%
%
%
%
\section{Results for hadron-nucleus cross sections}\label{data_results}
Total cross sections as well as their statistical and systematic
errors were determined for each dataset separately. In order to
calculate average total cross sections and average systematic 
errors, we use weighted means. We present the error contributions,
the data averaging method and the final results.
\subsection{Measurement errors}\label{meas_errors}
\subsubsection{The statistical error}\label{stat}
The dominant error contribution is the statistical error,
which is governed by the statistical uncertainty of the 
partial cross section $\sigma_{\rm part}(<|t_{\rm min}|)$, 
used in the ex\-tra\-po\-lation. Further statistical error 
contributions, originating in other terms of the error 
propagated formula~(\ref{xextrap}), are negligible. 
The statistical errors for each measurement are presented 
in table~\ref{tab:result}.
\subsubsection{Systematic errors}\label{syst}
In this section we briefly describe the systematic errors 
found during the data analysis. Table~\ref{tab:error_sizes}
gives an overview of the average sizes of these errors
as well as the rate correction and the contaminant 
correction.\\[0.5cm]
\noindent
Systematic error of the extrapolation $\errextr$ \\[0.15cm] \indent
A significant error contribution is the systematic 
error in the extrapolation of partial cross sections. This 
error takes into account the RMS-spread (root mean square) 
in the extrapolated total cross sections with respect to the 
extrapolated total cross section at $t_{\rm min} = -0.007\, 
\mbox{GeV$^2$/c$^2$}$, when $t_{\rm min}$ is varied from 
$-0.004\, \mbox{GeV$^2$/c$^2$}$ to $-0.01\, 
\mbox{GeV$^2$/c$^2$}$.\\[0.5cm]
Cut on the BTRD signal spectrum $\errbtrd$ \\[0.15cm] \indent
Although contaminant and rate corrections are applied for
each specific cut on the BTRD signal spectrum, we still
observe a variation of the cross section when varying the 
cut on the TRD plane count by $\pm$~1~unit around its 
nominal value. Therefore, we calculate a systematic error, 
which is the maximum spread in the cross sections found in 
the cut variation.\\[0.5cm]
Spill to spill fluctuations $\errfluc$ \\[0.15cm] \indent
Here, we compare the statistical error in
$\sigma_{\rm part}(<0.01\, \mbox{GeV$^2$/c$^2$})$, which we
calculate by~(\ref{error_part}) with the error in 
$\sigma_{\rm part}(<0.01\, \mbox{GeV$^2$/c$^2$})$ calculated 
from the experimentally observed RMS-spread of rate corrected 
transmission ratios per spill. The difference in these errors 
accounts for remaining non statistical spill to spill 
fluctuations.\\[0.5cm]
Systematic error of the rate correction $\errrate$ \\[0.15cm] \indent
This error takes into account the error arising from different 
functional attempts to describe the rate effect presented in 
section~\ref{rate_correction}. Its value is given by the maximum 
spread of the $\Delta_{\rm T0,k}$ with respect to their average 
value $\Delta_{\rm T0}$.\\[0.5cm]
Systematic error of the contaminant correction $\errcont$ \\[0.15cm] \indent
This systematic error accounts for the uncertainty in the 
fit parameters of the four-fold binomial 
distribution~(\ref{btrdfit}) and for the uncertainty in the 
contaminant fractions for $\Sigma^+$, $\Xi^-$ and K$^-$.\\[0.5cm]
Uncertainty of the target density $\errtgt$ \\[0.15cm] \indent
The target densities were measured several times, using a 
pycnometer and a buoyancy method. Laboratory studies showed
systematic discrepancies in the density measurement, which 
are included in the density errors shown in table~\ref{tab:targets}. 
These errors are propagated to an error contribution to the
total cross sections, which are on a 0.1\% 
level.\\
\par
\renewcommand{\arraystretch}{1.4}
\begin{table}[htbp]
  \begin{center}
    \scriptsize
    \begin{tabular}{|c|c|c|c|c|c|c|c|c|c|}
    \hline
                &         & \multicolumn{6}{|c|}{Systematic errors} &
                            \multicolumn{2}{|c|}{Corrections} \\
    cross       & $\plab$   & $\errextr$  & $\errbtrd$ & $\errfluc$ & $\errrate$ & 
    $\errcont$  & $\errtgt$ & $\ratecorr$ & $\contcorr$ \\
    section     & [GeV/c] &  [mb]   &  [mb]   &  [mb]   &  [mb]   & 
    [mb]     & [mb] &  [mb] & [mb] \\
    \hline
    $\totpbe$   & 536  & 0.91 & 0.70 &  0.25 & 0.35 & 0.06 & 0.30 & -1.24 & 0.62  \\
    \hline
    $\totsimbe$ & 638  & 1.20 & 0.49 &  0.04 & 0.10 & 0.07 & 0.27 & -0.93 & 0.65  \\
    \hline
    $\totpimbe$ & 638  & 0.50 & 0.17 &  0.22 & 0.05 & 0.61 & 0.21 & -0.79 & 1.00  \\
    \hline
    $\totpc$    & 457  & 0.90 & 2.11 &  0.54 & 0.38 & 0.09 & 0.47 & 11.22 & 0.91  \\
    \hline
    $\totpc$    & 490  & 1.81 & 1.53 &  0.68 & 1.15 & 0.10 & 0.47 & -3.87 & 0.86  \\
    \hline
    $\totsimc$  & 598  & 1.57 & 1.92 &  1.21 & 1.18 & 0.13 & 0.43 & -6.42 & 1.12  \\
    \hline
    $\totpimc$  & 591  & 1.30 & 1.40 &  1.50 & 0.95 & 0.63 & 0.33 & -3.11 & 1.03  \\
    \hline
    $\totsimpol$ & 589 & 2.10 & 2.55 &  0.69 & 0.16 & 0.16 & 0.30 & 3.67  & 1.44  \\ 
    \hline
    $\totpimpol$ & 585 & 1.26 & 0.96 &  0.54 & 0.12 & 0.75 & 0.23 & 2.90  & 1.21  \\
    \hline
    $\totsimcu$  & 609 & 163  & 41   &  76   & 41   & 0.33 & 1.23 & -754  & 3.1   \\
    \hline
    $\totpimcu$  & 608 & 85   & 52   &  78   & 36   & 2.99 & 1.03 & -649  & 4.7   \\
    \hline
    \end{tabular}
    \normalsize
    \vspace*{0.8cm}
    \caption{Average sizes of systematic errors and corrections. For explanation of 
             symbols see text of sections~\ref{t0ratecorr}, \ref{cont_corr} and
             \ref{meas_errors}.}
    \label{tab:error_sizes}
  \end{center}
\end{table}
\renewcommand{\arraystretch}{1.0}
\noindent
\subsection{Data-averaging and results on hadron-nucleus
cross sections}\label{averaging}
\subsubsection{The average total cross section}
Total cross-section results $\sigma_{\rm tot,i}$, obtained from 
i~=~1, ..., N data sets, are combined to an average total cross 
section $\overline{\sigma}_{\rm tot}$ with a statistical error 
$\delta^{\rm stat} \overline{\sigma}_{\rm tot}$ and an average 
systematic error $\delta^{\rm syst}\overline{\sigma}_{\rm tot}$.
The results are shown in table~\ref{tab:result}.\\ \indent
We average the total cross-sections $\sigma_{\rm tot,i}$ that
correspond to a specific measurement using the weighted 
mean:
\begin{equation}
  \label{averageing}
  \overline{\sigma}_{\rm tot} \ = \ 
  \frac{ \sum\limits_{\rm i = 1}^{\rm N} \omega_{\rm i} \ \sigma_{\rm tot,i} 
  }{ 
  \sum\limits_{\rm i = 1}^{\rm N} \omega_{\rm i} } 
  \ \ \ , \ \ \ 
  \omega_{\rm i} \ = \ \frac{1}{ (\delta_{\rm i}^{\rm stat})^2  \ + \ 
  \sum\limits_{\rm j = 1}^{\rm M} (\delta_{\rm j,i}^{\rm syst})^2   } \ .
\end{equation}
The weight $\omega_{\rm i}$ includes the statistical error 
($\delta_{\rm i}^{\rm stat}$) of data set~i and all 
j~=~1, ..., M systematic errors $\delta_{\rm j,i}^{\rm syst}$ 
described in section~\ref{stat} and~\ref{syst}.\\
As the statistical error is supposed to decrease, when adding more
data to the evaluation, we calculate the statistical error in the 
averaged total cross section by:
\begin{equation}
  \delta^{\rm stat} \overline{\sigma}_{\rm tot} \ = \ 
  \sqrt{ 1 / \sum\limits_{\rm i = 1}^{\rm N} \frac{1}{ 
  (\delta^{\rm stat}_{\rm i})^2  } } \ .
\end{equation}
In assigning a systematic error to an average total cross section
$\overline{\sigma}_{\rm tot}$ we assume that the systematic 
errors of the single measurements can be just averaged. Therefore,
we quote as average systematic error:
\begin{equation}
  \delta^{\rm syst}\overline{\sigma}_{\rm tot} \ = \
  \sqrt{\frac{ \sum\limits_{\rm i = 1}^{\rm N} \omega_{\rm i} 
  \left[  
  \sum\limits_{\rm j = 1}^{\rm M} (\delta^{\rm syst}_{\rm j,i})^2  
  \right]}{\sum\limits_{\rm i = 1}^{\rm N} \omega_{\rm i} }} \ ,
\end{equation}
which accounts for the weights $\omega_{\rm i}$ used in
(\ref{averageing}).\\ \indent
Further, we quote a total error 
$\delta^{\rm tot}\overline{\sigma}_{\rm tot}$
of the average total cross section, which is 
calculated from:
\begin{equation}
  \delta^{\rm tot}\overline{\sigma}_{\rm tot} \ = \ \sqrt{
  \left( \delta^{\rm stat}\overline{\sigma}_{\rm tot} \right)^2 \ + \ 
  \left( \delta^{\rm syst}\overline{\sigma}_{\rm tot} \right)^2       } \ .
\end{equation}
\renewcommand{\arraystretch}{1.4}
\begin{table}[htbp]
  \begin{center}
    \small
    \begin{tabular}{|c|c|c|c|c|c|}
    \hline
    cross &  $\plab$ &   $\overline{\sigma}_{\rm tot}$ &
    $\delta^{\rm stat} \overline{\sigma}_{\rm tot}$ &  $\delta^{\rm syst} 
     \overline{\sigma}_{\rm tot}$ & 
    $\delta^{\rm tot} \overline{\sigma}_{\rm tot}$ \\
    section                      & [GeV/c] & [mb] & [mb]   & [mb]   & [mb]   \\
    \hline
    $\sigma_{\rm tot}(\rm pBe)$          & 536     & 268.6   & $\pm$ 0.7 & $\pm$ 1.3 & $\pm$ 1.5 \\
    \hline
    $\sigma_{\rm tot}(\Sigma^- \rm Be)$   & 638     & 249.1   & $\pm$ 0.9 & $\pm$ 1.3 & $\pm$ 1.6 \\
    \hline
    $\sigma_{\rm tot}(\pi^- \rm Be)$      & 638     & 188.7   & $\pm$ 0.8 & $\pm$ 0.9 & $\pm$ 1.2 \\
    \hline
    $\sigma_{\rm tot}(\rm pC)$           & 457     & 333.6   & $\pm$ 3.1 & $\pm$ 2.4 & $\pm$ 3.9 \\
    \hline
    $\sigma_{\rm tot}(\rm pC)$           & 490     & 335.4   & $\pm$ 3.6 & $\pm$ 2.9 & $\pm$ 4.6 \\
    \hline
    $\sigma_{\rm tot}(\Sigma^-\rm C)$    & 598     & 308.9   & $\pm$ 2.1 & $\pm$ 3.8 & $\pm$ 4.3 \\
    \hline
    $\sigma_{\rm tot}(\pi^-\rm C)$       & 591     & 234.1   & $\pm$ 1.5 & $\pm$ 3.1 & $\pm$ 3.5 \\
    \hline
    $\sigma_{\rm tot}(\Sigma^-\rm CH_2)$ & 589     & 376.4   & $\pm$ 2.0 & $\pm$ 4.1 & $\pm$ 4.5 \\ 
    \hline
    $\sigma_{\rm tot}(\pi^-\rm CH_2)$    & 585     & 286.1   & $\pm$ 1.3 & $\pm$ 2.0 & $\pm$ 2.4 \\
    \hline
    $\sigma_{\rm tot}(\Sigma^-\rm Cu)$   & 609     & 1232    & $\pm$ 133 & $\pm$ 192  & $\pm$ 233 \\
    \hline
    $\sigma_{\rm tot}(\pi^-\rm Cu)$      & 608     & 1032    & $\pm$ 77  & $\pm$ 162  & $\pm$ 179 \\
    \hline
    \end{tabular}
    \normalsize
    \vspace*{0.8cm}
    \caption{Results for nuclear total cross sections. For explanation of 
             symbols see text of section~\ref{averaging}.}
    \label{tab:result}
  \end{center}
\end{table}
\renewcommand{\arraystretch}{1.0}
\subsection{Comparison to existing data on hadron-nucleus total cross 
sections}\label{data_comp1}
A literature survey showed that experimental data on 
hadron-nucleus total cross sections for charged projectiles
at high energies is extremely scarce. Information for 
proton-nucleus and $\pi^-$-nucleus total cross sections is 
only provided by the references~\cite{bell66,sch80,sch79} and
displayed together with our results in the 
figures~\ref{fig:pimcu}, \ref{fig:glau_probe}, 
\ref{fig:glau_proca}, \ref{fig:glau_pimbe} and 
\ref{fig:glau_pimca}. No data were found for $\Sigma^-$-nucleus 
total cross sections.
\subsubsection{Comparison of nucleon-nucleus total cross sections}
\begin{figure}[htbp]
  \begin{center}
  \leavevmode
  \epsfxsize=\hsize
  \epsfbox{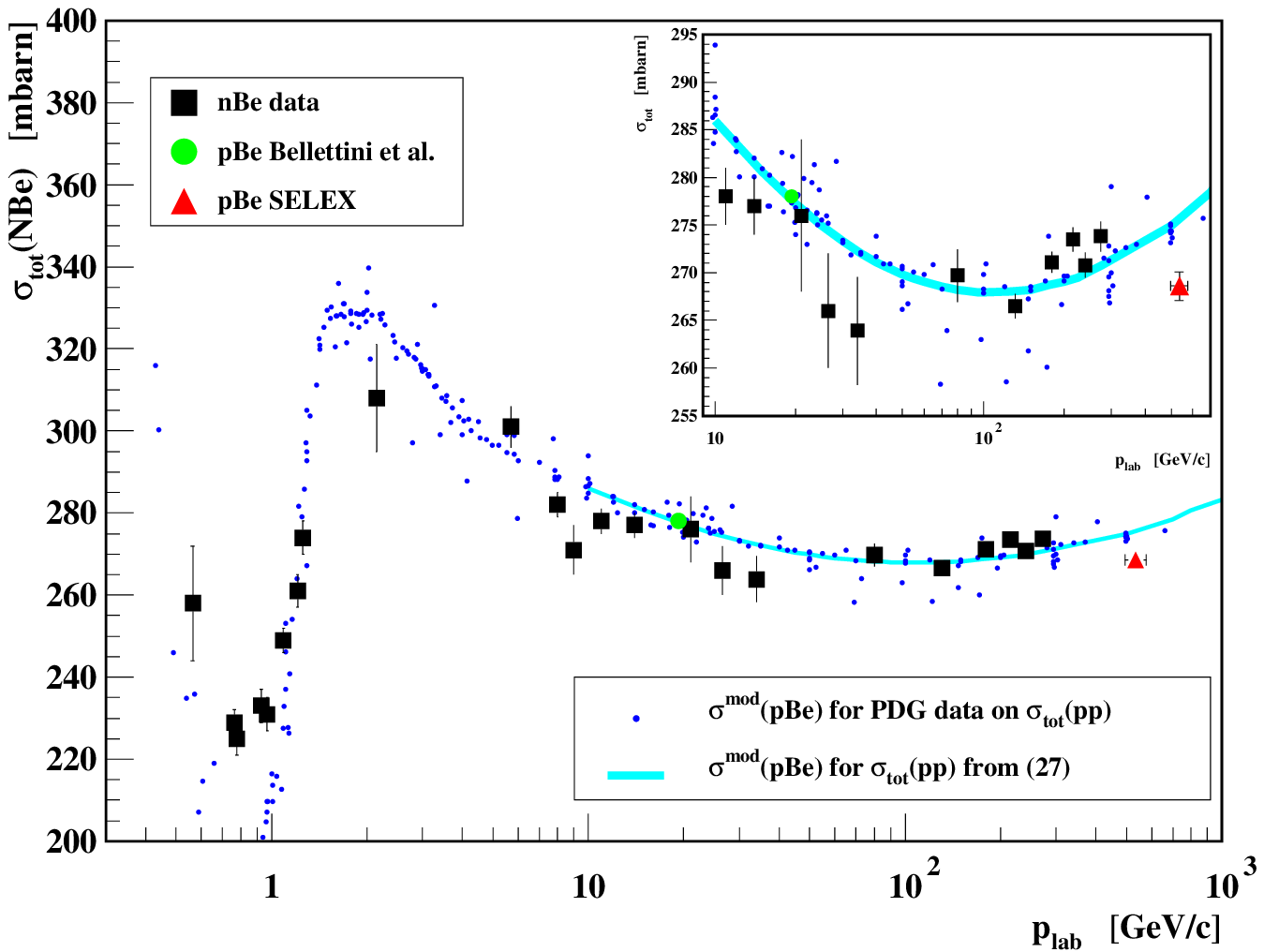}
  \end{center}
  \caption{Summary of experiment data on
           $\totnbe$ from~\cite{murt75,land73,engl70,bab74} 
           and on $\totpbe$ from~\cite{bell66} and SELEX.
           Overlaid are results from the model calculation
           (see section~\ref{model_calc}).}
  \label{fig:glau_probe}
\end{figure}

\begin{figure}[htbp]
  \begin{center}
  \leavevmode
  \epsfxsize=\hsize
  \epsfbox{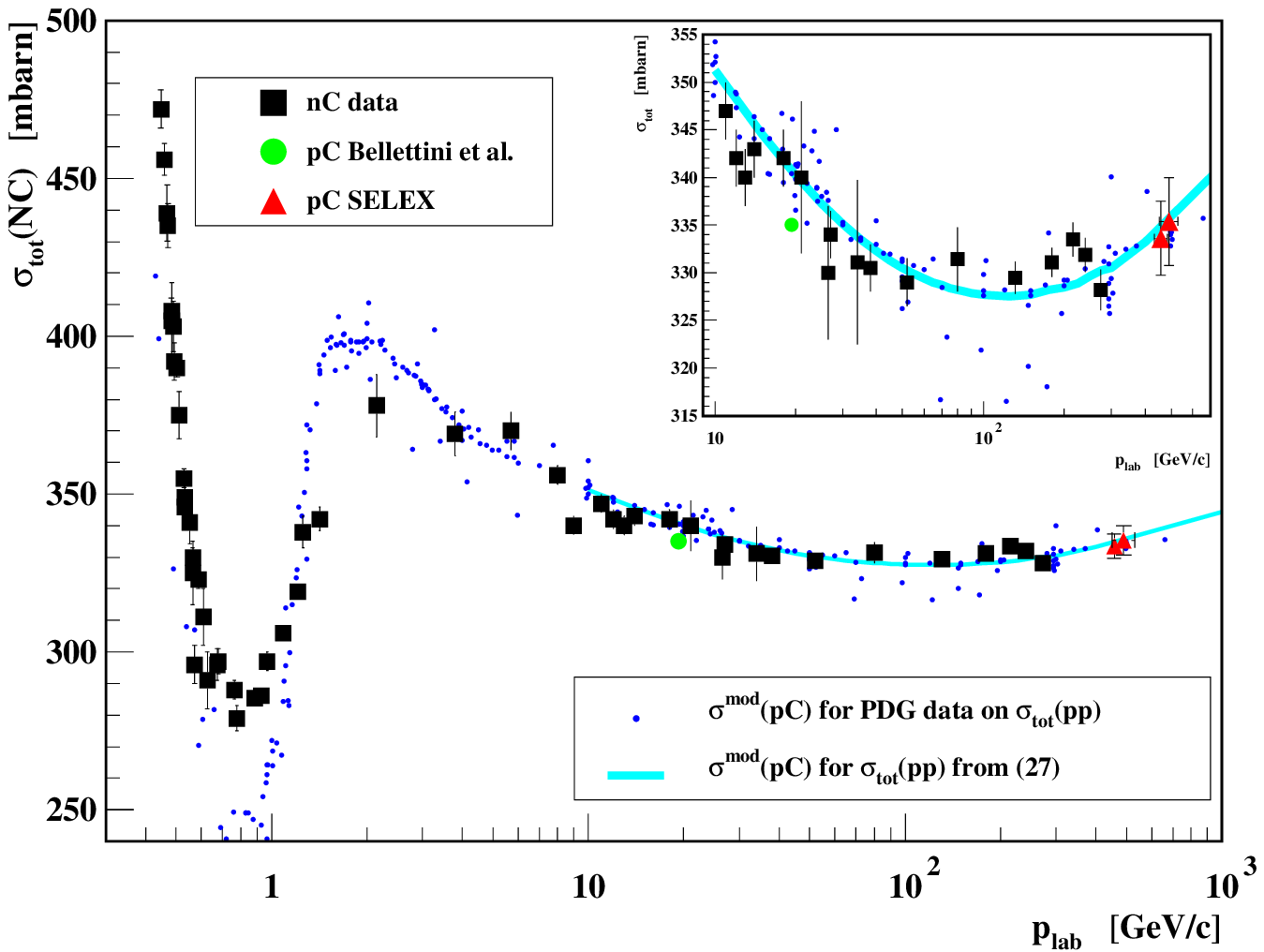}
  \end{center}
  \caption{Summary of experiment data on
           $\totnc$ from~\cite{murt75,land73,engl70,bab74} 
           and on $\totpc$ from~\cite{bell66} and SELEX.
           Overlaid are results from the model calculation
           (see section~\ref{model_calc}).}
  \label{fig:glau_proca}
\end{figure}

In the figures~\ref{fig:glau_probe} and \ref{fig:glau_proca} we
display a compilation of proton-nucleus and neutron-nucleus cross 
sections extracted from~\cite{murt75,land73,engl70,bab74} together
with our results. As can be seen, the proton-nucleus cross sections 
of~\cite{bell66} at $\plab$~$=$~$20\, \mbox{GeV/c}$ and the 
\noindent
neutron-nucleus cross sections are close by. For this reason we assume 
that differences between neutron-nucleus and proton-nucleus cross sections are 
negligibly small above $20\, \mbox{GeV/c}$. This allows a comparison of our
proton-nucleus cross sections with corresponding neutron-nucleus cross-section 
data available at much higher energy.\\ \indent
Comparing our results with neutron-nucleus cross sections at
131~--~$273\, \mbox{GeV/c}$ (data of~\cite{murt75}) shows 
that our measurements join the sequence of these data points. 
Averaging the neutron-beryllium total cross sections in this
\noindent
momentum range results in 271.0~$\pm$~$0.6\, 
\mbox{mbarn}$, which is close to our proton-beryllium cross section at 
$536\, \mbox{GeV/c}$ of 268.6~$\pm$~$1.5\, \mbox{mbarn}$. A similar 
calculation for the neutron-carbon cross section gives a mean value of 
331.0~$\pm$~$0.8\, \mbox{mbarn}$, which is close to our measurements of 
the proton-carbon cross section around $457\, \mbox{GeV/c}$ of 
333.6~$\pm$~$3.9\, \mbox{mbarn}$.
\subsection{Comparison of $\pi^-$-nucleus total cross sections}
\begin{figure}[htbp]
  \begin{center}
  \leavevmode
  \epsfxsize=\hsize
  \epsfbox{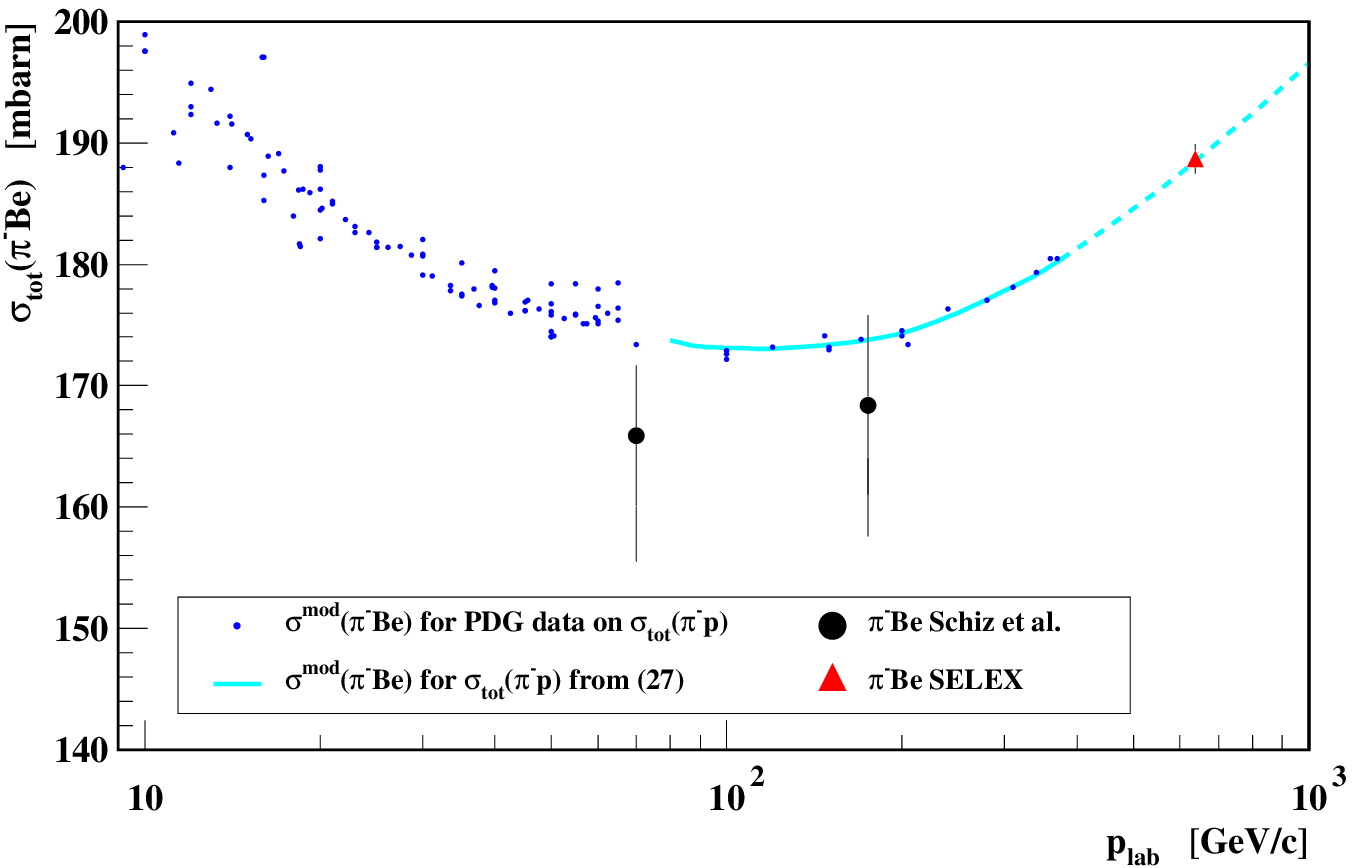}
  \end{center}
  \caption{Summary of experiment data on
           $\totpimbe$ from~\cite{sch80} and SELEX.
           Overlaid are results from the model calculation
           (see section~\ref{model_calc}).}
  \label{fig:glau_pimbe}
\end{figure}

\begin{figure}[htbp]
  \begin{center}
  \leavevmode
  \epsfxsize=\hsize
  \epsfbox{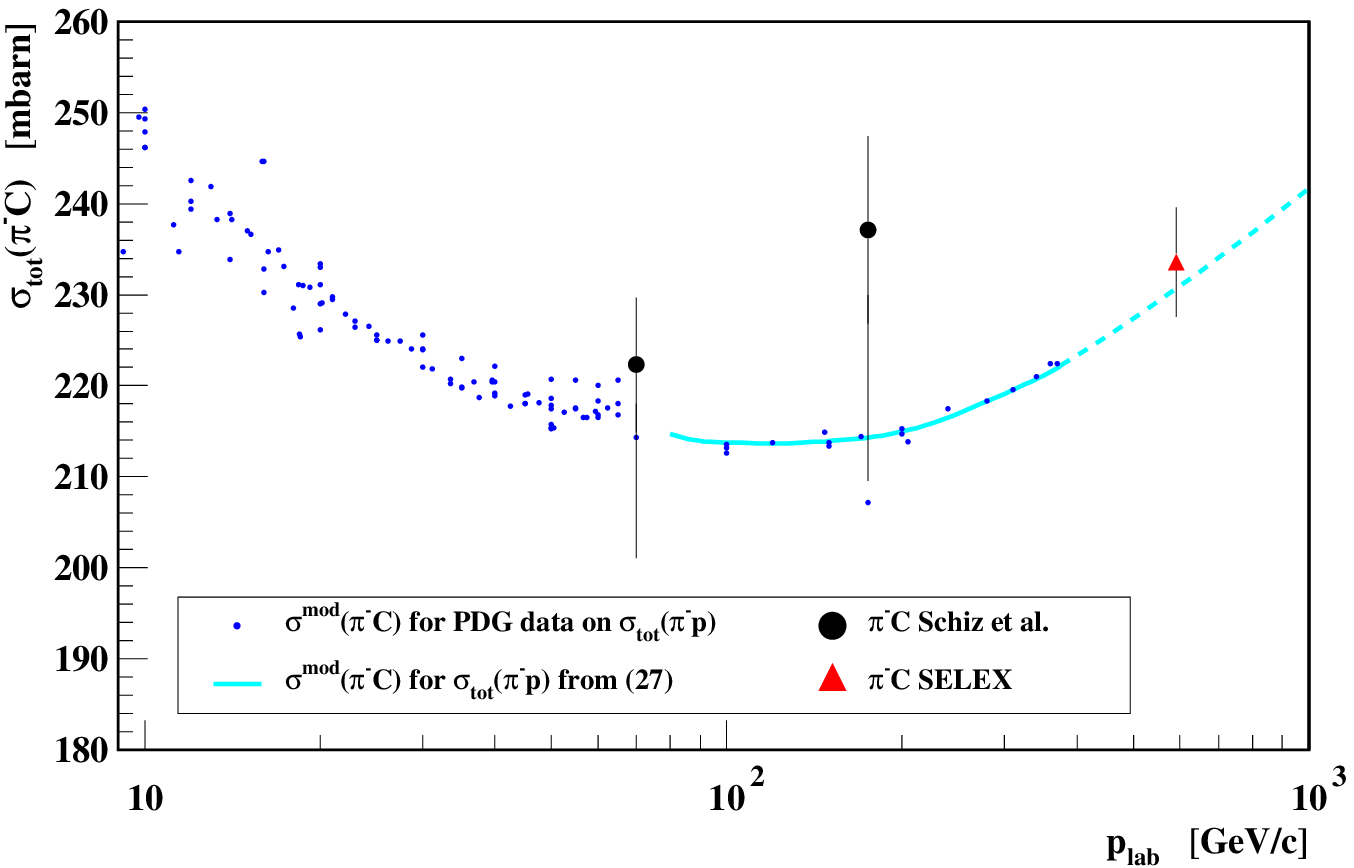}
  \caption{Summary of experiment data on
           $\totpimc$ from~\cite{sch80} and SELEX.
           Overlaid are results from the model calculation
           (see section~\ref{model_calc}).}
  \label{fig:glau_pimca}
  \end{center}
\end{figure}

\begin{figure}[htbp]
  \begin{center}
  \leavevmode
  \epsfxsize=\hsize
  \epsfbox{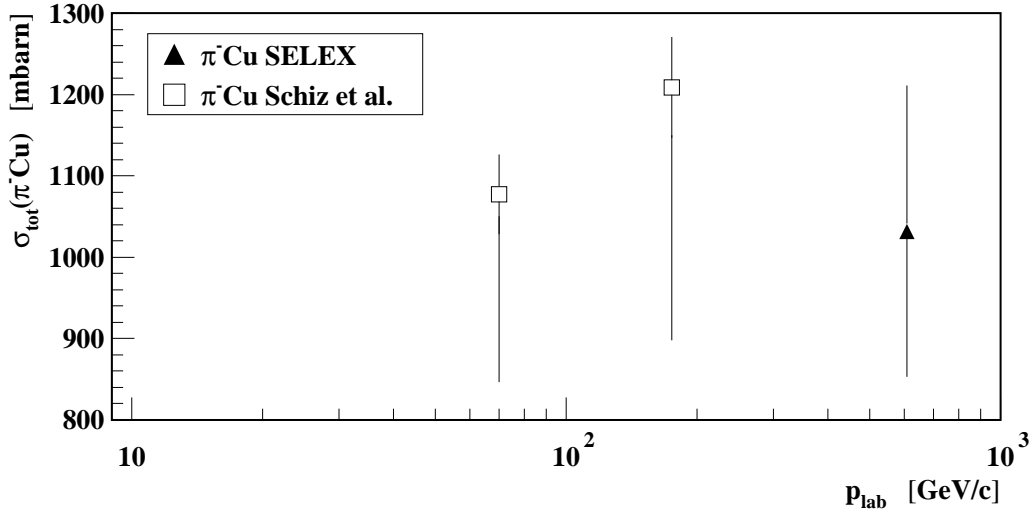}
  \end{center}
  \caption{Results for $\sigma_{\rm tot}(\pi^-\rm Cu)$ from
           \cite{sch80} and SELEX.}
  \label{fig:pimcu}
\end{figure}

High-energy data for $\sigma_{\rm tot}(\pi^-\rm Be)$, 
$\sigma_{\rm tot}(\pi^-\rm C)$ and $\sigma_{\rm tot}(\pi^-\rm Cu)$ that 
were determined using a transmission technique are presented in
the thesis of A. Schiz~\cite{sch79}. Unfortunately, the statistical
errors quoted for the $\pi^-$A total cross-sections are too large and
miss further corrections. Therefore, we present in figure~\ref{fig:pimcu},
\ref{fig:glau_pimbe} and \ref{fig:glau_pimca} ``normalization factors'' 
for $\pi^-$-nucleus scattering taken from~\cite{sch80}. These factors are
\noindent
\noindent
based on the thesis~\cite{sch79}, have the meaning of a total $\pi^-$-nucleus
cross section, but result from fits to $\pi^-$-nucleus elastic scattering 
data, and have smaller error.\\ \indent
The figures~\ref{fig:glau_pimbe},~\ref{fig:glau_pimca} and ~\ref{fig:pimcu} 
show that the SELEX results for $\pi^-$Be, 
$\pi^-$C and $\pi^-$Cu cross sections join the displayed normalization 
factors. However, further and more precise data would allow a more 
detailed comparison.
%
%
%
%
%
%
%
%
%
%
\section{Model description of hadron-nucleus cross 
sections}\label{model_calc}
In this section, we introduce a model calculation 
for hadron-nucleus cross sections and show how well it 
describes the data.
\subsubsection{The Glauber model and the inelastic screening 
correction}\label{gla_kar}
As shown in~\cite{murt75}, the Glauber model~\cite{glau59,fra72} 
including an inelastic screening correction~\cite{kar73}, is very 
precise in describing neutron-nucleus cross sections at high energy. 
The Glauber model accounts for the elastic screening effect in 
nuclei via multiple elastic scattering between the incident 
hadron $h$ and the nucleons $N$. As mentioned in~\cite{murt75}, nuclear 
total cross sections calculated by the Glauber model exceed experimental 
data. This is compensated by taking into account the inelastic screening
correction described in~\cite{kar73}. It accounts for inelastic reactions 
$h$~+~$N$~$\rightarrow$~$N$~+~$X$, which produce an inelastic screening 
effect. Consequently, a model cross section $\sigma_{\rm tot}^{\rm mod}$
comprises two parts:
\begin{equation}
\sigma_{\rm tot}^{\rm mod}(\rm A,\tothn ) \ = \ 
\sigma_{\rm tot}^{\rm Gla}(\rm A,\tothn ) \ - \
\Delta\sigma^{\rm Kar} \ .
\end{equation}
These are a Glauber model cross section 
$\sigma_{\rm tot}^{\rm Gla}(\rm A,\sigma_{\rm tot}(\rm hN))$
and an inelastic screening correction $\Delta\sigma^{\rm Kar}$.\\[0.5cm]
\underline{The Glauber model cross section}\\[0.15cm] \indent
According to~\cite{fra72}, $\sigma_{\rm tot}^{\rm Gla}(\rm A,\tothn )$ 
can be calculated by:
\begin{eqnarray}
 \label{glauform}
 \! \! \! \! \! \! \!
 \sigma_{\rm tot}^{\rm Gla}(\rm hA) & = & 4\pi \Re e \left\{
 \int_0^{\infty} 1 \ - \ \left[
 1 \ - \ \frac{(1-i\rho'_{\rm hN})}{2} \tothn \rm T(b) 
 \right]^{\rm A} \rm b db
 \right\}  \\
 \nonumber
 {\rm T(b)} & = & \frac{1}{2 \pi} \int_{\rm o}^{\infty} 
 \rm J_{\rm o}(qb) e^{-B_{\rm hN} \frac{\rm q^2}{2}} \rm  S(q) q dq \ \ \ , \ \ \  
 {\rm S(q)} \ = \ \frac{4 \pi}{\rm q} \int_{\rm o}^{\infty} \rm r \sin{\rm (qr)} 
 \tilde{\rho} \rm (r) dr \ .
\end{eqnarray}
Here $\rho'_{\rm hN}$ is the real to the imaginary part of the elastic
scattering amplitude in the forward direction observed in hadron-nucleon
elastic scattering and b is the impact parameter. $B_{\rm hN}$ is the 
hadronic slope in hadron-nucleon elastic scattering and J$_{\rm o}$ is 
a Bessel function of 0'th order. The nuclear density $\tilde{\rho} \rm (r)$ 
is normalized as:
\begin{equation}
 4 \pi \int_{0}^{\infty} \tilde{\rho}\rm (r) \rm r^2 dr \ = \ 1 \ .
\end{equation}
\underline{The inelastic screening correction}\\[0.15cm] \indent
The inelastic screening correction $\Delta\sigma^{Kar}$, originally
formulated in~\cite{kar73} for proton-nucleus reactions, is 
generalized by:
\begin{eqnarray}
  \label{karform}
  \Delta\sigma^{\rm Kar} & = &
  4 \pi \int_{\rm o}^{\infty} \int_{\rm (m_{\rm p} + 
  \rm m_{ \pi})^2}^{(\sqrt{ \rm s} - \rm m_{\rm p})^2} \! \!
  \left( \frac{\rm d^2 \sigma}{\rm dtdM^2} \right)_{\rm t = 0} \! \! \! \! \! \! \rm
  e^{-\frac{1}{2} \sigma_{\rm tot}\rm (hN) A
  \tilde{\rm T}\rm (b)} \left| \rm  F(q_L,\vec{\rm b})  \right|^2 \rm dM^2 d^2b  \\
  \nonumber
  \tilde{\rm T}(\vec{\rm b}) & = & \int_{- \infty}^{+ \infty} 
  \tilde{\rho}(\vec{\rm b}\rm ,z) dz
  \ \ \ , \ \ \ 
  {\rm F(q_L,b)} \ = \ \rm A \int_{- \infty}^{+ \infty} \tilde{\rho}\rm (b,z) 
  e^{i \rm q_{\rm L} z} \rm   dz \\
  \nonumber
  \rm q_{\rm L} & = & (\rm M^2 \ - \ m_{\rm p}^2)\frac{m_{\rm p}}{\rm s} \ \ .
\end{eqnarray}
Here m$_{\rm p}$ is the proton mass and m$_{\pi}$ is the pion mass. The double
differential cross section d$^2\sigma$/dtdM$^2$ describes the inelastic 
reaction $h$~+~$N$~$\rightarrow$~$N$~+~$X$ of the incident hadron $h$ with a 
nucleon $N$, where the resulting final state $X$ has an invariant mass 
squared of M$^2$.
\subsubsection{Input parameters for the total cross-section model}
Model input parameters are $\sigma_{\rm tot}\rm (hN)$, 
$\rho'_{\rm hN}$, $B_{\rm hN}$, $\tilde{\rho}$(r) and 
$\rm (d^2\sigma/\rm dt dM^2)|_{\rm t=0}$. All of them are
extracted from experimental data with $N$~=~p.
\newpage
\noindent
\underline{Input parameter $\sigma_{\rm tot}\rm (hN)$}\\[0.15cm] \indent
To be able to access $\sigma_{\rm tot}\rm (hN)$ for special
ranges of center of mass energies $\sqrt{\rm s}$, data on pp
and $\pi^-$p total cross sections, taken from~\cite{pdg98}, 
were fit to a smooth function:
\begin{equation}
 \label{hp_fitfunction}
 \sigma_{\rm tot}\rm (hp,s) \ = \ 
 \frac{\rm a_{\rm o}}{\rm s^{\rm a_{\rm 1}}} \ + \ 
 \rm a_{\rm 2} \log^2 ( s ) \ .
\end{equation}
The fit-parameters $a_{\rm i}$, their errors and the
validity range of each parameterization, are shown in 
table~\ref{tab:hn_fit}. The result of each parameterization is
in mbarn, when using s in GeV$^2$.\\
\par
\begin{figure}[htbp]
  \begin{center}
  \leavevmode
  \epsfxsize=\hsize
  \epsfbox{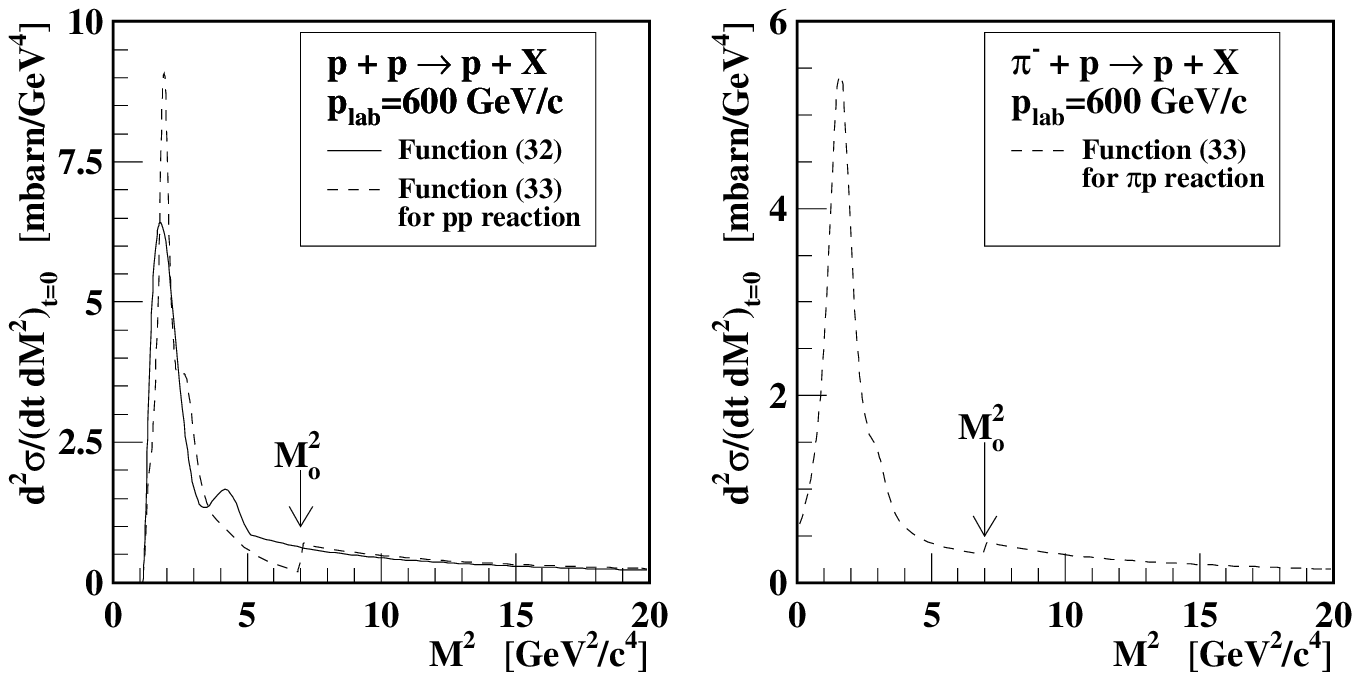}
  \end{center}
  \caption{The parameterizations~(\ref{doubdiff_murthy}) 
           and~(\ref{doubdiff_dakhno}) evaluated for 
           $\plab$ = $600\, \mbox{GeV/c}$.}
  \label{fig:inclusive}
\end{figure}

\renewcommand{\arraystretch}{1.4}
\begin{table}[htbp]
  \begin{center}
    \begin{tabular}{|c|c|c|c|c|}
    \hline
    reaction & a$_{\rm o}$      & a$_{\rm 1}$       & a$_{\rm 2}$ & momentum range \\
    \hline
    pp       & 49.51 $\pm$ 0.26 & 0.097 $\pm$ 0.002 & 0.314 $\pm$ 0.004 & 
    10 ... 3000 GeV/c \\
    $\pi^-$p & 55.2  $\pm$ 7.2  & 0.255  $\pm$ 0.032  & 0.346  $\pm$ 0.020  & 
    80 ... 380  GeV/c \\
    \hline
    \end{tabular}
    \vspace*{0.8cm}
    \caption{Fit-Parameters and validity range of the total cross-section 
             parameterizations.}
    \label{tab:hn_fit}
  \end{center}
\end{table}
\renewcommand{\arraystretch}{1.0}
\noindent
\vspace*{0.1cm}\\
\underline{Input parameter $\rho'_{\rm hp}$}\\[0.15cm] \indent
We parameterize $\rho'_{\rm pp}( \plab )$ 
and $\rho'_{\pi^- \rm p}( \plab )$, using data 
on $\rho'_{\rm pp}$ from \cite{burq83,gross78,beznogikh72,vorobyov72,foley67,fajardo81,jenni77,breedon89,amos83,amaldi77,amos85,akopin77,amirkhanov73} and data on 
$\rho'_{\pi^- \rm p}$ from \cite{burq83,fajardo81,foley69,apokin76},
\begin{figure}[htbp]
  \begin{center}
  \leavevmode
  \epsfxsize=\hsize
  \epsfbox{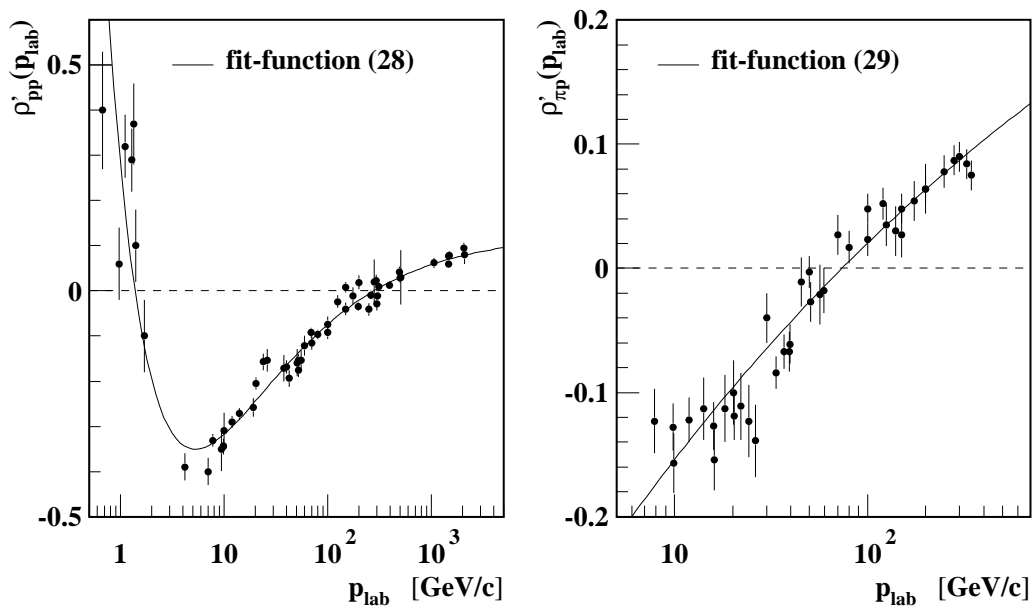}
  \end{center}
  \caption{Our parameterizations for $\rho'_{\rm pp}( \plab )$ and 
           $\rho'_{\pi^- \rm p}( \plab )$ together with experimental data
           from \cite{burq83,gross78,beznogikh72,vorobyov72,foley67,fajardo81,jenni77,breedon89,amos83,amaldi77,amos85,akopin77,amirkhanov73}.}
  \label{fig:retoim}
\end{figure}
\noindent
assuming that $\rho'$ reaches a constant value when $\plab$ 
goes to infinity. Our fits are
\begin{eqnarray}
  \rho'_{\rm pp}( \plab )    \ & = & \ + \frac{6.8}{  \plab^{0.742}}   \ - \
                                     \frac{6.6}{ \plab^{0.599}}   \ + \
                                     0.124 \\
  \nonumber
  & & \  \textnormal{for 0.8~GeV/c~$<$~$\plab$~$<$~2100~GeV/c} \\[0.5cm]
  \rho'_{\pi^- \rm p}( \plab ) \ & = & \ - \frac{0.92}{ \plab^{0.54}}   \ + \
                                   0.54 \\
  \nonumber
  &   & \  \textnormal{for 8.0~GeV/c~$<$~$\plab$~$<$~345~GeV/c} \ ,
\end{eqnarray}
\noindent
where $\plab$ is in GeV/$c$. Figure~\ref{fig:retoim} displays these 
fit-functions together with all data points included in the 
fit.\\[0.5cm]
\underline{Input parameter $B_{\rm hp}$}\\[0.15cm] \indent
For the hadronic slope parameters $B_{\rm pp}$ and $B_{\pi^- \rm p}$ 
we take the parameterizations presented in~\cite{burq82}:
\renewcommand{\arraystretch}{1.4}
\begin{eqnarray}
\! \! \! \! \! \!
 \rm B_{\rm pp}( \plab )   & = & \left\{ \begin{array}{l@{\quad}crll}
 \rm B_{\rm pp,1} & = & 11.13 & - \frac{6.21}{\sqrt{ \plab }} + 0.30 \log 
 \{ \plab \} & \ \ ; \ \ \rm q^2 = 0.02 \\
 \rm B_{\rm pp,2} & = &  9.26 & - \frac{4.94}{\sqrt{ \plab }} + 0.28 \log 
 \{ \plab \} & \ \ ; \ \ \rm q^2 = 0.20 \\
 \rm B_{\rm pp,3} & = &  9.67 & - \frac{7.51}{\sqrt{ \plab }} + 0.10 \log 
 \{ \plab \} & \ \ ; \ \ \rm q^2 = 0.40
                               \end{array} \right. \\
                   &  & \nonumber \\[0.2cm]
 \rm B_{\pi p}( \plab ) & = & \left\{ \begin{array}{l@{\quad}crll }
 \rm B_{\pi \rm p,1} & = &  9.11 & + \frac{0.65}{\sqrt{ \plab }} + 0.29 \log 
 \{ \plab \} & \ \ ; \ \ \rm q^2 = 0.02 \\
 \rm B_{\pi \rm p,2} & = &  6.95 & + \frac{0.65}{\sqrt{ \plab }} + 0.27 \log 
 \{ \plab \} & \ \ ; \ \ \rm q^2 = 0.20 \\
 \rm B_{\pi \rm p,3} & = &  6.13 & + \frac{0.65}{\sqrt{ \plab }} + 0.25 \log 
 \{ \plab \} & \ \ ; \ \ \rm q^2 = 0.40 \ .
                               \end{array} \right.
\end{eqnarray}
\renewcommand{\arraystretch}{1.0}
\noindent
Here, q$^2$ is in units of GeV$^2$/$c^2$. These parameterizations
are linearly interpolated to account for the dependency of 
B$_{\rm hN}$ on both $\plab$ and q$^2$.\\[0.5cm]
\underline{Input parameter (d$^2\sigma$/dtdM$^2)|_{\rm t=0}$}\\[0.15cm] \indent
To calculate the inelastic screening correction $\Delta\sigma^{\rm Kar}$,
we use the parameterization of (d$^2\sigma$/dtdM$^2)|_{\rm t=0}$ 
for the process p~+~p~$\rightarrow$~p~+~$X$, given in~\cite{murt75}:
\renewcommand{\arraystretch}{1.1}
\begin{eqnarray}
  \label{doubdiff_murthy}
  \left( \frac{\rm d^2 \sigma}{\rm dt dM^2} \right)_{\rm t=0} \! \! \! & = & 
  \left\{ \begin{array}{l@{\quad\quad}l }
  26.470 (\rm M^2-1.17)  \ - \ 35.969(\rm M^2-1.17)^2 \ + & \\
  18.470(\rm M^2-1.17)^3 \ - \  4.143(\rm M^2-1.17)^4 \ + & \\
  0.341(\rm M^2-1.17)^5  \ \ \textnormal{for 1.17 $<$ M$^2$ $<$ 5 GeV$^2$/$c^2$} & \\[0.2cm]
  \frac{4.4}{\rm M^2}    \ \ \ \ \ \ \ \ \ \ \ \ \ \ \ \ \ \ \ \ \ \
                           \textnormal{for M$^2$ $>$ 5 GeV$^2$/$c^2$} &  \ .    
          \end{array} \right. 
\end{eqnarray}
\renewcommand{\arraystretch}{1.0}
\noindent
In addition, we also use more recent parameterizations for 
$(\rm d^2\sigma/\rm dt dM^2)|_{\rm t=0}$ to describe the processes 
p~+~p~$\rightarrow$~p~+~$X$ and $\pi$~+~p~$\rightarrow$~p~+~$X$, 
which are presented in~\cite{dakhno83} and are based on calculations 
of triple-Regge diagrams in~\cite{kazarinov76}. For M$^2$~$\le$~M$_{\rm o}^2$, 
these parameterizations consist of a background term and a sum of 
non-energy-dependent resonance terms. 
In case M$^2$~$>$~M$_{\rm o}^2$ the parameterizations 
consist of a sum over contributions from triple-Regge diagrams:
\renewcommand{\arraystretch}{1.8}
\begin{equation}
\label{doubdiff_dakhno}
\! \! \! \! \!
\left( \frac{\rm d^2 \sigma}{\rm dt dM^2} \right)_{\rm t=0} \ = \ 
\left\{ \begin{array}{l@{\quad\quad}l }
  \sum\limits_{\rm i} \frac{\rm a_{\rm i}}{\rm (M^2-M_{\rm i}^2)^2 + 
  \Gamma_{\rm i}} \ + \ 
  \frac{\rm c_{\rm f} \rm (M^2-M_{\rm min}^2 )}{\rm (M^2-M_{\rm min}^2)^2 + 
  \rm d_{\rm f}} \ \ , 
  \ \rm M^2 \le \rm M_{\rm o}^2  & \\
  \sum\limits_{\rm k} \rm V_{\rm k} \left(\frac{\rm M^2}{\rm s} 
  \right)^{\alpha_{\rm k(0)}-
  \beta_{\rm k(0)}-\beta'_{\rm k(0)}}
  \frac{1}{\rm s^{\rm 2-\alpha_{\rm k(0)}}} \ \ , \ \rm M^2 > \rm M_{\rm o}^2 & \ . 
             \end{array} \right.
\end{equation}
\renewcommand{\arraystretch}{1.0}
\noindent
Instead of displaying the large amount of parameters for
equation~(\ref{doubdiff_dakhno}), which are taken from
calculations in~\cite{kazarinov76}, we display the 
parameterizations~(\ref{doubdiff_murthy}) and~(\ref{doubdiff_dakhno})
in figure~\ref{fig:inclusive}.\\
\par
\noindent
Compared to~(\ref{doubdiff_dakhno}), parameterization~(\ref{doubdiff_murthy}) 
has no s-dependence. Further, parameterization~(\ref{doubdiff_dakhno}) is not 
continuous and the resonance sizes are quite different for 
p~+~p~$\rightarrow$~p~+~$X$.\\[0.5cm]
\underline{Input parameter $\tilde{\rho}(\rm r)$}\\[0.15cm] \indent
In the calculations, we use density distributions $\tilde{\rho}(\rm r)$ 
that are based on the harmonic-oscillator model:
\begin{equation}
  \tilde{\rho}(\rm r) \ = \ \rho_{\rm o} \left[ 1 \ + \ \tilde{\alpha} 
  \left( \frac{\rm r}{a_{\rm rad}} 
  \right )^2 \right ] \rm e^{- \left( \frac{\rm r}{a_{\rm rad}} \right )^2 } \ .
\end{equation}
This offers the possibility to calculate some integrals in an analytic 
way and gives a better description of the (charge-) density
distribution for light nuclei than a standard two-parameter Fermi 
parameterization.\\
\par
\renewcommand{\arraystretch}{1.0}
\begin{table}[htbp]
  \begin{center}
    \begin{tabular}{|c|c|c|c|c|}
    \hline
              & \multicolumn{2}{|c|}{data from~\cite{jag74}}             & 
                fit result using~(\ref{doubdiff_murthy})                 &
                fit result using~(\ref{doubdiff_dakhno})                 \\
    nucleus   & \multicolumn{2}{|c|}{e-A scattering}                      &
                in $\sigma_{\rm tot}^{\rm mod}(\rm NA,\rm a_{\rm rad})$   & 
                in $\sigma_{\rm tot}^{\rm mod}(\rm NA,\rm a_{\rm rad})$   \\
              &  $\tilde{\alpha}$    & a$_{\rm rad}$ [fm]                 & 
                 a$_{\rm rad}$ [fm]  & a$_{\rm rad}$ [fm]                 \\
    \hline
    beryllium &  0.611               & 1.791    &   1.89981 & 2.02914 \\
    carbon    &  1.067               & 1.687    &   1.79247 & 1.89277 \\
    \hline
    \end{tabular}
    \vspace*{0.8cm}
    \caption{Parameters of the density distribution $\tilde{\rho}(\rm r)$ 
             from electron-nucleus elastic scattering~\cite{jag74} and the radius
             parameters resulting from a fit of
             $\sigma_{\rm tot}^{\rm mod}(\rm a_{\rm rad}, \plab )$
             to nA cross-section data in the momentum 
             range 10~--~$273\, \mbox{GeV/c}$.}
    \label{tab:HOpar}
  \end{center}
\end{table}
\renewcommand{\arraystretch}{1.0}
\noindent
As reported in~\cite{murt75}, we also find that the model does not provide 
a good description of neutron-nucleus total cross sections if one
uses both $\tilde{\alpha}$ and a$_{\rm rad}$ from electron-scattering 
data~\cite{jag74}. Therefore, we used $\tilde{\alpha}$ values 
from~\cite{jag74} and adjusted the radius parameter a$_{\rm rad}$, such that 
the model cross section $\sigma_{\rm tot}^{\rm mod}(\rm NA,\rm a_{\rm rad})$ 
gives a best description of nA-cross section data in the momentum range 
10~--~$273\, \mbox{GeV/c}$. Adjusting of a$_{\rm rad}$ was done for each 
nucleus and for each of the parameterizations~(\ref{doubdiff_murthy})
and~(\ref{doubdiff_dakhno}) separately. Table~\ref{tab:HOpar} gives a summary 
of the density parameters.
\subsubsection{Results of the model calculations}\label{quality_check}
\noindent
\underline{Results for nucleon-nucleus model cross sections}\\[0.15cm] 
To show the quality of our model calculation after
adjusting the nuclear density parameter a$_{\rm rad}$, we evaluated 
the total cross sections $\modnbe$ and $\modnc$ using
function~(\ref{doubdiff_murthy}). This was done for data on 
$\totpp$ taken from~\cite{pdg98} and for values on $\totpp$ 
resulting from our fit~(\ref{hp_fitfunction}). The calculations 
were done at many different values of $\plab$ to show the 
behavior over the entire high momentum region. Scatter in the 
model calculations (observed when experimental data on 
$\totpp$ are used) demonstrate the sensitivity of the model 
to small changes in $\totpp$.\\ \indent
Summaries of calculation and data are shown in 
figure~\ref{fig:glau_probe} and~\ref{fig:glau_proca}. They show
that the calculations reflect quite well the cross-section data
for $\plab$~$>$~$5\, \mbox{GeV/c}$. The nBe data of~\cite{murt75}
in the range 131~--~$273\, \mbox{GeV/c}$ suggest a rise of the
nBe cross section with energy that is also indicated by the model calculation. 
Our data point does not show any rise for pBe. In the case of nC cross 
sections our measurements join both data at lower energy and calculation 
very nicely.\\[0.5cm]
\noindent
\underline{Results for $\pi^-$-nucleus model cross sections}\\[0.15cm]
We evaluated the cross sections $\modpimbe$ and 
$\modpimc$ using function~(\ref{doubdiff_dakhno}) and
the corresponding nuclear density parameter a$_{\rm rad}$,
which was determined by a fit of the model cross section to 
neutron-nucleus data. All further 
input parameters are specific for $\pi^-$p-reactions.
The calculations were done for data on $\totpimp$
taken from~\cite{pdg98} and for values from 
function~(\ref{hp_fitfunction}).\\ \indent
Results are shown in figure~\ref{fig:glau_pimbe} 
and~\ref{fig:glau_pimca} together with data for $\pi^-$-nucleus 
total cross sections from~\cite{sch80} and the SELEX 
experiment. The figures show that the calculations match
our measurements quite well and agree within errors with
lower-energy data from~\cite{sch80}.
%
%
%
%
%
%
%
\section{Results for hadron-nucleon cross sections}\label{data_nn}
The hadron-nucleon cross sections $\sigma_{\rm tot}(\Sigma^- \rm N)$ and 
$\sigma_{\rm tot}(\pi^- \rm N)$ were first determined by a CH$_2$~--~C
method. As this method provides hadron-nucleon cross sections only with 
a precision on the order of 10\%, we improved the precision using a method
which takes advantage of the more precise hadron-nucleus cross-section 
ratios.
\subsection{Hadron-nucleon cross sections using a CH$_2$~--~C 
difference method}\label{method1}
The hadron-nucleon cross sections $\sigma_{\rm tot}(\Sigma^- \rm N)$ 
and $\sigma_{\rm tot}(\pi^- \rm N)$ can be deduced from corresponding cross 
sections measured on carbon and polyethylene by:
\begin{equation}
\sigma_{\rm tot}(\rm hN) \ = \ \frac{1}{2} \left[ 
\sigma_{\rm tot}(\rm hCH_2) \ - \ \sigma_{\rm tot}(\rm hC)
\right] \ ,
\end{equation}
where h denotes the incident hadron. Results obtained by this method 
are presented in table~\ref{tab:result_had_nuc}. The quoted errors are
calculated from the total errors in the hadron-nucleus cross sections
given in table~\ref{tab:result}.
\subsection{Hadron-nucleon cross sections deduced from
            hadron-nucleus cross sections}\label{method2}
In a second approach, we deduce hadron-nucleon cross sections
from ratios of measured hadron-nucleus cross sections. To
motivate the method, we first derive empirical relations
between hadron-nucleon and hadron-nucleus cross section 
ratios, which we then refine using the model calculation 
described in section~\ref{model_calc}.\\ \indent
To derive empirical relations between hadron-nucleon and 
hadron-nucleus cross section ratios we use data on 
hadron-nucleon cross sections around $137\, \mbox{GeV/c}$ 
from~\cite{bia81,pdg98}, and obtain the hadron-nucleon 
cross-section ratios:
\begin{equation}
\label{hn_ratios}
  \frac{\totpimp}{\totpp} \ \approx \  0.635 \pm 0.006,  \ \ \ \ 
  \frac{\totsimp}{\totpp} \ \approx \  0.901 \pm 0.012 \ .
\end{equation}
Next, we build nuclear cross-section ratios using
our measurements for the $\Sigma^-$A, $\pi^-$A and pA 
cross sections from table~\ref{tab:result}.\\ \indent
Our pA cross sections were measured at lower laboratory 
momentum than the corresponding $\Sigma^-$A or $\pi^-$A cross 
sections. To correct for this, we scale the pA cross 
sections by a factor k$_{\rm scale}$ before building the
cross-section ratio. The scale factor takes into account 
the growth of the pA cross section from the laboratory 
momentum where it was measured to the larger laboratory 
momentum of the corresponding $\Sigma^-$A or $\pi^-$A cross 
section. Scaling factors are calculated using the model 
described in section~\ref{model_calc}. They are displayed 
together with the nuclear cross-section ratios in 
table~\ref{tab:xratios}.\\
\par
\renewcommand{\arraystretch}{1.1}
\begin{table}[htbp]
  \begin{center}
    \begin{tabular}{|lc|c|c|}
    \hline
    \multicolumn{2}{|c|}{scaled cross-section ratio} & 
    $\plab$ [GeV/c] & k$_{\rm scale}$  \\ 
    \hline
    $\totpimbe / \totpbe$ \hspace*{-0.4cm}  
    & = 0.698 $\pm$ 0.006   & 640           & 1.0058  \\
    $\totpimc  / \totpc$  \hspace*{-0.4cm}  
    & = 0.695 $\pm$ 0.014   & 590           & 1.0036  \\
    $\totsimbe / \totpbe$ \hspace*{-0.4cm}  
    & = 0.922 $\pm$ 0.008   & 640           & 1.0058  \\
    $\totsimc  / \totpc$  \hspace*{-0.4cm}  
    & = 0.917 $\pm$ 0.018   & 590           & 1.0040  \\
    \hline
    \end{tabular}
    \vspace*{0.8cm}
    \caption{Nuclear cross-section ratios. The pA-cross section is 
             scaled by k$_{\rm scale}$ to account for the discrepancy 
             in laboratory momenta of the cross sections used in
             the ratio.}
    \label{tab:xratios}
  \end{center}
\end{table}
\renewcommand{\arraystretch}{1.0}
\noindent
The nuclear ratios show that the $\pi^-$A cross sections are about 
0.7 times and the $\Sigma^-$A cross sections are about 0.92 times 
smaller than the pA cross section.\\ \indent
To get a first relation between hadron-nucleon and hadron-nucleus
cross sections, we ignore the weak energy dependence of the
cross-section ratios. Calculating the ratios of hadron-nucleon
to hadron-nucleus cross-section ratios using the above data 
gives the results presented in table~\ref{tab:double_ratio}.
\par
\renewcommand{\arraystretch}{1.8}
\begin{table}[htbp]
  \begin{center}
    \begin{tabular}{|c|c|c|c|}
    \hline
    double ratio & result & double ratio & result \\
    \hline
    $\frac{\totsimp / \totpp}{\totsimbe / \totpbe}$ & 0.977 $\pm$ 0.016 &
    $\frac{\totpimp / \totpp}{\totpimbe / \totpbe}$ & 0.910 $\pm$ 0.012 \\
    $\frac{\totsimp / \totpp}{\totsimc  / \totpc}$  & 0.983 $\pm$ 0.023 &
    $\frac{\totpimp / \totpp}{\totpimc / \totpc}$   & 0.915 $\pm$ 0.020 \\
    \hline
    average ($\kappa$)                              & 0.980 $\pm$ 0.014 &
    average ($\kappa$)                              & 0.913 $\pm$ 0.012 \\
    \hline
    \end{tabular}
    \vspace*{0.8cm}
    \caption{Ratios of the hadronic cross-section ratios at
             137 GeV/c and the nuclear cross-section ratios
             around 600 GeV/c.}
    \label{tab:double_ratio}
  \end{center}
\end{table}
\renewcommand{\arraystretch}{1.0}
\noindent
The double ratios show a small but significant deviation from one
especially for ratios involving $\pi^-$ cross sections. From this
empirical observation it follows that a hadron-nucleon cross section 
$\sigma_{\rm tot}(hN)$ can be approximately derived from the pp 
cross section and a hadron-nucleus cross-section ratio using
the relation:
\begin{equation}
  \label{appox_xsec}
  \sigma_{\rm tot}(hN) \ \approx \kappa \ \times \totpp \times 
  \left( 
  \frac{\sigma_{\rm tot}(hA)}{\sigma_{\rm tot}(pA)} 
  \right) \ ,
\end{equation}
where $\kappa$ is a parameter specific for the 
cross section ratio (compare with table~\ref{tab:double_ratio}). 
If we set $\kappa$~=~1 for simplicity, we see that the 
precision of (\ref{appox_xsec}) is about 10\%. The precision is
improved by adequate adjusting of $\kappa$.\\ \indent
Unfortunately we cannot empirically derive $\kappa$ from experimental 
cross sections for laboratory momenta around 600 GeV/c as necessary 
cross-section data is missing. Thus, as we want to deduce hadron-nucleon 
cross sections from nuclear cross-section ratios with best precision, 
we improve the relation between hadron-nucleon and hadron-nucleus 
cross sections using the total cross-section model that was introduced in 
section~\ref{model_calc}.\\ \indent
The idea of the model-based ratio method is the following: 
Rewriting~(\ref{appox_xsec}) yields the following relation
between the experimental hadron-nucleus and the model based
hadron-nucleus cross-section ratios.
\begin{equation}
  \label{ratio_method_formula}
  \underbrace{\frac{\overline{\sigma}_{\rm tot}(hA)}{\overline{\sigma}_{\rm tot}
  (\rm pA)}}_{\textnormal{experimental}} \ = \
  \underbrace{\frac{\sigma_{\rm tot}^{\rm mod}(\rm A,
  \sigma_{\rm tot}(\rm hN))}{\sigma_{\rm tot}^{\rm mod}(\rm A,
  \sigma_{\rm tot}(\rm pN))}}_{\textnormal{theory + $\sigma_{\rm tot}$-data}} \ .
\end{equation}
Taking the ratio of model based quantities reduces the effect of
uncertainties in the cross-section model. Further, as data for 
$\totpp$ is most precise and exists over a large energy range, 
it is convenient to use proton-nucleus cross sections in the 
denominator. As the energy dependence of the pp cross
section is known at SELEX energies and the model is adjusted to 
describe NA cross sections for $\plab > 10$ GeV/c, the energy 
dependence of $\tothp$, which we want to determine, is 
taken into account through the energy dependence of the
pp cross section. \\ \indent
To deduce the cross section $\tothn$ from the measured nuclear
cross-section ratio, we fix $\totpn$ (= $\totpp$) first and 
calculate the denominator 
\linebreak
$\modpa$ by taking $\totpp$ from 
parameterization~(\ref{hp_fitfunction}) evaluated at the laboratory 
momentum of the nuclear cross-section ratio as given in 
table~\ref{tab:xratios}. Iterating with respect to the model input 
parameter $\tothn$ until the model based total cross-section ratio 
in~(\ref{ratio_method_formula}) equals the experimental one,
yields the desired hadron-nucleon cross section. At SELEX energy 
we interpret the result $\tothn$ identical to $\tothp$. 
\subsubsection{Results for $\sigma_{\rm tot}(\Sigma^- \rm N)$ and
$\sigma_{\rm tot}(\pi^- \rm N)$ using the ratio method}
Results of the ratio method are presented in 
table~\ref{tab:result_had_nuc} together with the results
from the CH$_2$~--~C method. The errors of hadron-nucleon 
cross sections resulting from the ratio method include both
the error in the measured nuclear cross-section ratio and 
model uncertainties. Model uncertainties are taken into account 
by adding the error of a model cross-section ratio in quadrature 
to the error of the corresponding experimental cross-section ratio 
given in table~\ref{tab:xratios}. The error in the model 
cross-section ratio is derived from the discrepancy between model 
and measured cross sections observed for pA and $\pi^-$A total 
cross sections. Typical sizes of these discrepancies are 
shown in table~\ref{tab:discrepancy}.\\
\par
\renewcommand{\arraystretch}{1.0}
\begin{table}[htbp]
  \begin{center}
    \begin{tabular}{|c|c|c|c|c|}
    \hline
             & measured cross & calculated    & cross-section & nominal       \\
    reaction & section $\times$ $k_{\rm scale}$ 
    & cross section & difference    & $\plab $ \\
             & [mbarn]       & [mbarn]       & [mbarn]       & [GeV/c]       \\
    \hline
    $\totpimbe$ &  188.7     & 188.8 & 0.1 & 640 \\
    $\totpimc$  &  234.1     & 231.4 & 2.7 & 590 \\
    $\totpbe$   &  270.2     & 277.0 & 6.8 & 640 \\
    $\totpc$    &  336.8     & 335.9 & 0.9 & 590 \\
    \hline
    \end{tabular}
    \vspace*{0.8cm}
    \caption{Discrepancy between model and measured total
             cross sections. The measured pA cross sections
             are scaled by $k_{\rm scale}$.}
    \label{tab:discrepancy}
  \end{center}
\end{table}
\renewcommand{\arraystretch}{1.0}
\noindent
Further, as two different parameterizations for 
$(\rm d^2\sigma/\rm dt dM^2)|_{\rm t=0}$ are available, we 
evaluate the ratio method for both, average the results
and include their difference in the error of the 
mean. \\ \indent
Finally, we want to mention that as little data exists for 
$\Sigma^-$ scattering, we insert in the computation of 
$\sigma_{\rm tot}^{\rm mod}(\Sigma^-\rm A)$ for 
B$_{\Sigma^- \rm N}$, $\rho'_{\Sigma^- \rm N}$ and 
$(\rm d^2\sigma/\rm dt dM^2)|_{\rm t=0}$, the parameterizations 
from pp-reactions.\\ 
\par
\renewcommand{\arraystretch}{1.2}
\begin{table}[htbp]
  \begin{center}
    \begin{tabular}{|c|c|c|c|}
    \hline
    method      & $\sigma_{\rm tot}(\Sigma^- \rm N)$     &  
                  $\sigma_{\rm tot}(\pi^- \rm N)$        & 
                  $\plab $                          \\
    description & [mbarn] & [mbarn] & [GeV/c] \\
    \hline
    difference method & 33.7 $\pm$ 3.1    &  26.0 $\pm$ 2.1  & 585    \\
    \hline
    ratio method, Be data
                & 37.4 $\pm$ 1.3    & 27.1  $\pm$ 1.5  & 640    \\
    \hline
    ratio method, C data
                & 37.0 $\pm$ 0.8    & 26.4  $\pm$ 1.3  & 595    \\
    \hline
    \hline
    total average  &  37.0 $\pm$ 0.7 & 26.6 $\pm$ 0.9  & 610  \\
    \hline
    \end{tabular}
    \vspace*{0.8cm}
    \caption{The total cross sections $\sigma_{\rm tot}(\Sigma^- \rm N)$ and
             $\sigma_{\rm tot}(\pi^- \rm N)$ resulting from all
             methods and their average.}
    \label{tab:result_had_nuc}
  \end{center}
\end{table}
\renewcommand{\arraystretch}{1.0}
\noindent
Comparing the hadron-nucleon cross sections of the ratio and the 
difference method, we find good agreement of the results with respect 
to their errors. As final result, we average the hadron-nucleon 
cross-section values from all methods. These total averages are 
presented in the last row of table~\ref{tab:result_had_nuc} together
with a corresponding averaged laboratory momentum.
\subsection{Comparison to models}\label{data_comp2}
\subsubsection{Comparisons for 
$\sigma_{\rm tot}(\pi^- \rm p)$}\label{comp_pi}
\begin{figure}[htbp]
  \begin{center}
  \leavevmode
  \epsfxsize=\hsize
  \epsfbox{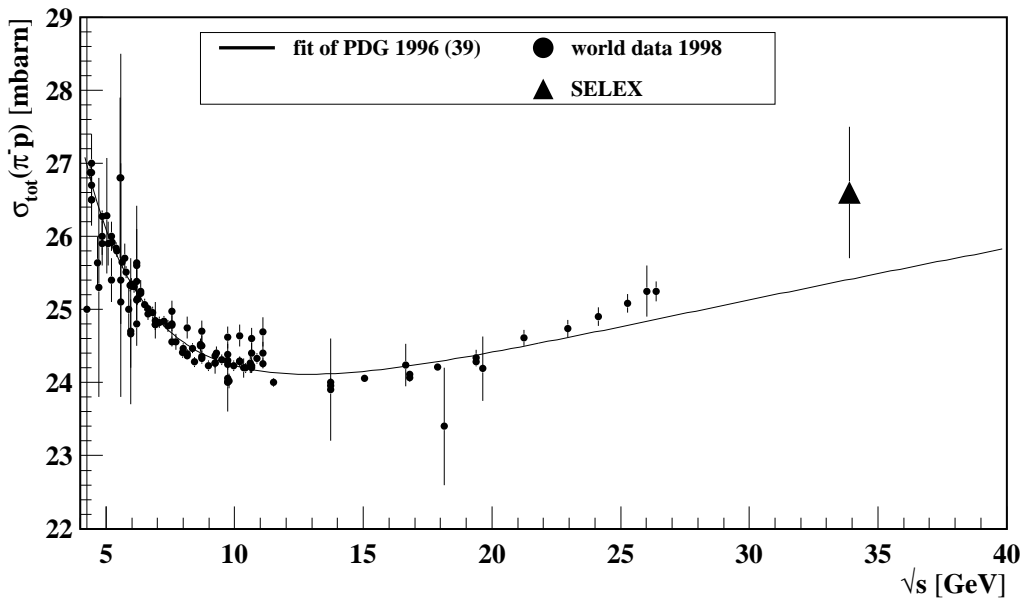}
  \end{center}
  \caption{Existing data for $\sigma_{\rm tot}(\pi^- \rm p)$ in comparison
           with our results and parameterization (\ref{pim_p_xsec_pdg1996})
           of the particle data group 1996.}
  \label{fig:pim_p_xsec}
\end{figure}

Most of the models and parameterizations for hadron-nucleon cross
sections exploit the interplay of 2 contributions. The Pomeron
contribution, which dominates asymptotics at high energies; and the 
Regge contribution, which is important at low and medium energies.
Many models (e.g.~\cite{don92,lipkin75}) describe the energy 
dependence of total cross sections quite well. Because of this, we 
simply display in figure~\ref{fig:pim_p_xsec} experimental data 
from~\cite{pdg98} and SELEX together with the parameterization for 
$\sigma_{\rm tot}(\pi^- \rm p,s)$:
\begin{eqnarray}
  \label{pim_p_xsec_pdg1996}
  \sigma_{\rm tot}(\pi^- \rm p,s) & = & 35.9 \rm  s^{-0.45} \ + \ 
                                        13.7 \rm  s^{+0.079} \\
  \nonumber
  & &  \textnormal{for $\plab \ >$ $10\, \mbox{GeV/c}$, 
  $\sigma_{\rm tot}$ in mbarn, s in GeV$^2$} \ ,
\end{eqnarray}
which was presented in the particle data book 1996~\cite{pdg96}. 
We want to mention that in the data files of the particle
\noindent
data group errors in the high precision data of A.S. 
Caroll et al.~\cite{car79} are not completely taken over
from the reference. Thus, we added the missing error
contributions (systematic error due to target density 
and extrapolation uncertainty) in quadrature.\\ \indent
We point out that so far the total cross section
$\sigma_{\rm tot}(\pi^- \rm p)$ has been measured only 
up to $\plab$~=~$370\, \mbox{GeV/c}$~\cite{car79}. 
Thus, the SELEX total average for $\sigma_{\rm tot}(\pi^- \rm N)$ 
at $610\, \mbox{GeV/c}$ is the first new measurement at higher 
laboratory momentum.\\ \indent
In figure~\ref{fig:pim_p_xsec} parameterization~(\ref{pim_p_xsec_pdg1996}) 
of the Particle Data Group, which uses a Pomeron intercept of~0.079, 
is overlaid to the data. The qualitative inspection 
of~(\ref{pim_p_xsec_pdg1996}) suggests that it is strongly weighted 
by the huge amount of low energy data points and does not sufficiently 
well take into account the very accurate data of~\cite{car79} at high
energy. Our result seems to strengthen the trend observed in
data of~\cite{car79}. This trend implies a faster rise of the
$\pi^-$p cross section with increasing energy than represented
by~(\ref{pim_p_xsec_pdg1996}). We just want to point out 
this observation, which may turn out to be in conflict
with the belief that the energy increase of hadronic
cross sections is universal. Further, we do not give any 
quantitative estimate of the Pomeron intercept for the $\pi^-$p 
cross section. Its value is correlated to the assumed Regge 
contribution at low energy and its determination 
requires a careful analysis of the data.
\subsubsection{Comparisons for $\sigma_{\rm tot}(\Sigma^- \rm p)$}
Up to now, data on the total cross section 
$\sigma_{\rm tot}(\Sigma^- \rm p)$ are still scarce. In the past, 
there have been only two hyperon-beam experiments~\cite{bia81,bad72} 
giving information about the behavior of 
$\sigma_{\rm tot}(\Sigma^- \rm p)$ in the momentum range 
\begin{figure}[htbp]
  \begin{center}
  \leavevmode
  \epsfxsize=\hsize
  \epsfbox{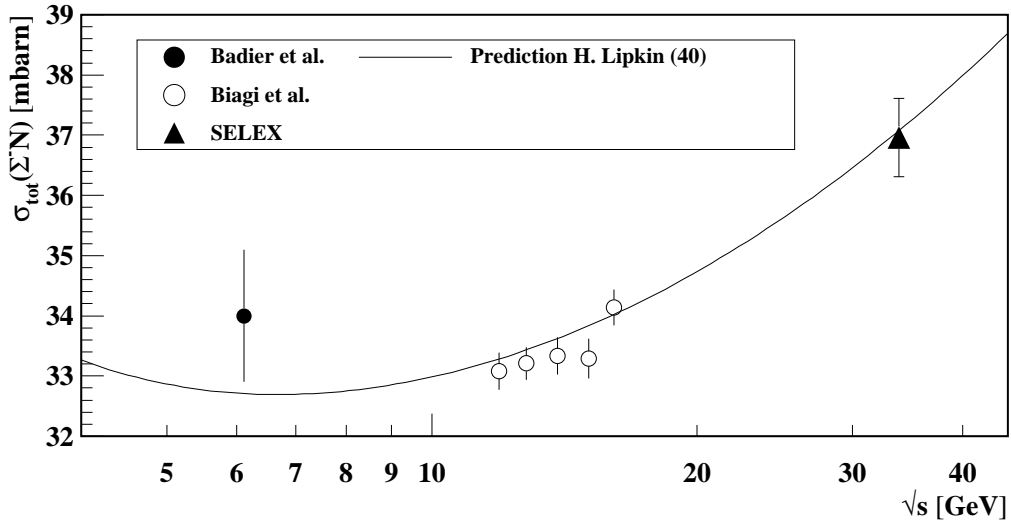}
  \end{center}
  \caption{Existing data for $\sigma_{\rm tot}(\Sigma^- \rm p)$ in comparison
           with our results and prediction~(\ref{simp_lipkin}).}
  \label{fig:sigma_p_xsec}
\end{figure}
\noindent
19~--~$136.9\, \mbox{GeV/c}$. The SELEX total average for
$\sigma_{\rm tot}(\Sigma^- \rm N)$ provides first new data at 
higher energy. Figure~\ref{fig:sigma_p_xsec} shows a compilation of 
data from previous experiments together with the SELEX result. Our 
measurement is $2.9\, \mbox{mbarn}$ larger than the data point at 
$136.9\, \mbox{GeV/c}$ from~\cite{bia81}. It shows the rise of 
$\sigma_{\rm tot}(\Sigma^- \rm p)$ with increasing beam 
energy.\\ \indent
Overlaid to the experiment data is the prediction for
$\sigma_{\rm tot}(\Sigma^- \rm p,\plab )$ from 
H. Lipkin (see~\cite{lipkin75}), which is given by:
\begin{eqnarray}
\label{simp_lipkin}
\sigma_{\rm tot}(\Sigma^- \rm p, \plab ) & = & 
  19.5 \left(\frac{ \plab }{20}\right)^{0.13} \ + \
  13.2 \left(\frac{ \plab }{20}\right)^{-0.2} \\
\nonumber
  & & \textnormal{for $\plab \ >$ $10\, \mbox{GeV/c}$, 
    $\sigma_{\rm tot}$ in mbarn, $\plab $ in GeV/c} \ .
\end{eqnarray}
The corresponding curve in figure~\ref{fig:sigma_p_xsec} shows good 
agreement between our measurement and this prediction.\\ \indent
It would be certainly desirable to find the Pomeron intercept for
the $\Sigma^-$p cross section. But the lack of low energy data 
does not allow any reasonable estimate of the intercept.
%
%
%
%
%
%
%
%
%
\section{Conclusions}
The SELEX collaboration has measured the total cross
sections $\totpimbe$, $\totpimc$, $\totpimcu$, $\totpimpol$, 
$\totsimbe$, $\totsimc$, $\totsimcu$,
\linebreak
$\totpbe$ and $\totpc$
in a broad momentum range around $600\, \mbox{GeV/c}$ using a transmission
method that was adapted to the specifics of the SELEX 
spectrometer. The accuracy of the results is within 0.6\% - 1.5\% 
for Be, C and CH$_2$ and about 17.5\% for Cu.\\ \indent
The ratios of hadron-nucleus cross sections for Be and C show that
$\pi^-$-nucleus cross sections are a about factor of 0.7 lower 
than corresponding proton-nucleus cross sections. Furthermore, we 
find that the $\Sigma^-$-nucleus cross sections are about a factor 
of 0.92 smaller than corresponding proton-nucleus cross 
sections.\\ \indent
We observe that the results for $\sigma_{\rm tot}(\rm pBe)$,
$\sigma_{\rm tot}(\rm pC)$, $\sigma_{\rm tot}(\pi^- \rm Be)$, 
$\sigma_{\rm tot}(\pi^- \rm C)$
and $\sigma_{\rm tot}(\pi^- \rm Cu)$ join smoothly corresponding cross-section 
data at lower energy. The good agreement 
of the proton-nucleus and the $\pi^-$-nucleus cross sections to Glauber 
model calculations which include an inelastic screening correction and 
one adjustable parameter in the density distribution, justify the deduction 
of $\sigma_{\rm tot}(\Sigma^- \rm p)$ and $\sigma_{\rm tot}(\pi^- \rm p)$ from 
the nuclear cross sections.\\ \indent
We deduced the hadron-nucleon cross sections $\sigma_{\rm tot}(\pi^- \rm N)$ 
and $\sigma_{\rm tot}(\Sigma^- \rm N)$, which we regard as $\sigma_{\rm tot}(\pi^- \rm p)$ 
and $\sigma_{\rm tot}(\Sigma^- \rm p)$, from our nuclear data using a CH$_2$~--~C 
difference and a model based ratio method. Results from the difference 
method have an accuracy of 8.1--9.2\%, while results from the ratio 
method have an accuracy of 2.2--5.5\%.\\ \indent
The total averages of all methods represent first new measurements
for $\sigma_{\rm tot}(\pi^- \rm p)$ and $\sigma_{\rm tot}(\Sigma^- \rm p)$ 
at higher energy. Our result for $\sigma_{\rm tot}(\Sigma^- \rm p)$ shows 
clearly a rise of this cross section with increasing beam energy, which agrees 
with the prediction of~\cite{lipkin75}.\\ \indent
Our result for $\sigma_{\rm tot}(\pi^- \rm p)$ joins nicely the high
energy data of~\cite{car79}. As mentioned in section~\ref{comp_pi},
the data of~\cite{car79} and our result may indicate a faster increase
of the $\pi^-\rm p$ cross section than predicted by the
parameterization given by the Particle Data Group in 1996.\\ \indent
An indication of a faster increase
of the $\pi^-\rm p$ cross section compared to the pp (and 
$\overline{\rm p}$p) one should be verified by a high statistic 
measurement using a $\pi^-$ beam and a hydrogen target to avoid 
systematic errors inherent to the method used in this experiment. 
In our opinion a measurement of the $\pi^-\rm p$ cross section 
at $600\, \mbox{GeV/c}$ or higher is the only experimentally
accessible case to test if the high energy behaviour
of the hadronic cross section can be different from the pp and
(and $\overline{\rm p}$p) one.
%
%
%
%
%
%
\section{Acknowledgments}
We are indebted to the Fermilab staff for the realization of this
experiment. We wish to thank F. Pearsall and 
D. Northacker for technical support. Further, we are grateful
to  T. Olzanovski, V. Mallinger, U. Schwan, R. Schwan, 
J. Zimmer and the mechanic workshop of the Max-Planck-Institute 
for nuclear physics at Heidelberg for support in organizing, 
machining and measuring target materials. We thank B. Kopeliovich 
and A.V. Tarasov for theoretical support and are very grateful to 
G.T. Garvey, J. Pochodzalla, H.W. Siebert and M. Zavertyaev for 
various discussions and contributions to the analysis. Last but 
not least it was a pleasure for us to 
contact H. Lipkin about model predictions.\\ \indent
This project was supported in part by the Bundesministerium f\"ur
Bildung, Wissenschaft, Forschung und Technologie, the Consejo
Nacional de Ciencia y Tecnolog\'{\i}a (CONACyT), the Conselho Nacional
de Desenvolvimento Cient\'{\i}fico e Tecnol\'{o}gico, the Fondo de Apoyo
a la Investigaci\'{o}n (UASLP), the Funda\c{c}\~{a}o de Amparo \`{a} 
Pequisa do Estado de S\~{a}o Paulo (FAPESP), the Israel Science Foundation 
founded by the Israel Academy of Sciences and Humanities, the Istituto 
Nazionale di Fisica Nucleare (INFN), the International Science Foundation 
(ISF), the National Science Foundation (Phy\#9602178), NATO (grant 
CR6.941058-1360/94), the Russian Academy of Science, the Russian Ministry 
of Science and Technology, the Turkish Scientific and Technological Research 
Board (T\"UB\.ITAK), the U.S. Department of Energy (DOE grant 
DE-FG02-91ER40664), and the U.S.-Israel Binational Science Foundation 
(BSF).
%
%
%
%

%
%
\end{document}